\documentclass[12pt]{article}

\usepackage{amsmath,amssymb,amsfonts,graphicx}
\usepackage{ucs}
\usepackage{bigints}
\usepackage{epstopdf}
\usepackage{color}
\usepackage[utf8]{inputenc}
\usepackage{amsthm}
\usepackage{fontenc}
\usepackage{tikz}
\RequirePackage{slashed}
\usepackage{cancel}
\usepackage[normalem]{ulem}

\usepackage{latexsym}
\usepackage{amssymb}
\usepackage{booktabs}
\usepackage{comment}
\usepackage{authblk}
\usepackage{float} 
\usepackage{cite}
\usepackage{xcolor}

\usepackage[a4paper,top=2cm,bottom=2cm,left=2cm,right=2cm]{geometry}

\usepackage{braket}

\newcommand{\bit}{\begin{itemize}}
\newcommand{\eit}{\end{itemize}}

\definecolor{mygreen}{rgb}{0.0, 0.5, 0.0}    
\newcounter{comment}

\newlength\savedwidth

\title{Two-current correlations 
in the pion in the Nambu and Jona-Lasinio model
}

\begin{document}

\author[1]{Aurore Courtoy}
\affil[1]{\small{ Instituto de F\'isica, Universidad Nacional Aut\'onoma de M\'exico\\
Apartado Postal 20-364, 01000 Ciudad de M\'exico, Mexico} }
\author[2]{Santiago Noguera}
\affil[2]{\small{ Departamento de F\`isica Te\`orica and IFIC, Centro Mixto Universidad de Valencia-CSIC,
E-46100 Burjassot (Valencia), Spain } }
\author[3]{Sergio Scopetta}
\affil[3]{ \small{Dipartimento di Fisica e Geologia, Universit\`a degli Studi di Perugia, and Istituto Nazionale 
di Fisica 
Nucleare,
Sezione di Perugia, Via A. Pascoli, I-06123, Perugia, Italy } }

\maketitle

\begin{abstract}
{We present an analysis of two-current correlations for the pion
%The non perturbative information encoded 
%in two current correlations { in the pion} is investigated 
in the Nambu--Jona-Lasinio model, with Pauli--Villars regularization.  
We provide explicit expressions in momentum space for two-current correlations corresponding
to the zeroth component of the vector Dirac bilinear in the quark vertices, which has been evaluated on the lattice.
The numerical results show a
remarkable qualitative agreement with recent lattice data.
 The factorization approximation into one-body currents is discussed.
}

\end{abstract}

\section{\label{sec:intro}Introduction}

Due to the high partonic densities reached, processes with more than two partons from the two colliding protons participating in the actual scattering ---the so called multiple parton interactions--- are likely to happen at the LHC~\cite{Bartalini:2017jkk}.
Double parton scattering (DPS), 
the simplest form of  multiple parton interactions,
involves two simultaneous hard collisions. It  has been indeed  observed at the LHC (see, e.g., Ref. \cite{Aad:2013bjm}).
The DPS cross section is expressed in terms of double parton distribution functions (dPDFs)~\cite{Paver:1982yp,Diehl1}.
The latter are related to the number density of two partons located at a given transverse separation in 
coordinate space with given longitudinal momentum fractions.
%The information encoded in  these distributions therefore complements the so-called hadron tomography, accessed through electromagnetic probes \comminlineAC{DY also gives access to TMDs and PDFs and probably GPDs, does it count as elm probe? Not sure}, 
%in terms of , generalized parton distributions (GPDs),
The information encoded in  these distributions therefore complements that accessed through electromagnetic probes in terms of  generalized parton distributions (GPDs)~\cite{Guidal:2013rya,Dupre:2016mai}, one of the flagships of the future Electron Ion Collider (EIC)~\cite{Aidala:2020mzt}.  
If measured, dPDFs would therefore represent 
a novel tool to study the three-dimensional hadron structure \cite{blok1,blok2,fabbro,ffe,Rinaldi:2018bsf}.
Indeed, they are 
sensitive 
to  two-parton correlations not accessible via one body 
distributions, {\it e.g.} 
Parton Distribution Functions (PDFs) and GPDs --see Ref. \cite{Bartalini:2017jkk,
Kasemets:2017vyh}
for a recent report.
Since they are non perturbative objects, dPDFs have
 been evaluated within models of the hadron structure ---estimates exist at low momentum scales
for the proton dPDFs~\cite{bag,noi2,noi1,kase,plb,JHEP2016,Traini:2016jru}. 
In order to match theoretical predictions with future experimental 
analyses,  the results of these  calculations are then evolved using the proper QCD evolution equations to reach the  high momentum scale  of the data. 
Developments on the evolution properties
of dPDFs 
can be found in, {\it e.g.}, 
\cite{Diehl:2017wew}.

Recently, the double PDFs of the pion have also attracted considerable attention.
A first estimate has been performed using light cone wave functions obtained within the AdS/QCD correspondence~\cite{Rinaldi:2018zng,
Rinaldi:2020ybv} ;
then a calculation 
within a Nambu--Jona-Lasinio (NJL) framework was presented in Ref.~\cite{jhep19}.
The latter represents the first evaluation of pion dPDFs in a field theoretical approach, which
allowed a systematical study of the different contributions to the pion observables. This exercise has been  effective to examine the validity of commonly used approximations for which dPDFs could be expressed
in terms of GPDs ---and were found to be violated, especially at low momentum scale and in the valence region.
These results have been recently confirmed in a similar
NJL framework by an independent calculation \cite{bar}.

Studies for the pion are especially important
in view of possible lattice calculations. This is  pertinent 
in particular
when a corresponding evaluation for  nucleon observables on the lattice  appear much more involved.
However, the communities studying parton distributions from global QCD analyses and lattice QCD calculations are collaborating 
towards a synergy to further improve our knowledge of PDFs~\cite{Lin:2017snn}, thanks  
 to the access --on the lattice-- to moments of PDFs as well as to the formalism of quasi-PDFs and its matching procedure to lightfront PDFs, {\it e.g.}~\cite{Ji:2013dva}.
This effort is expanding to all collinear PDFs, GPDs and Transverse momentum dependent parton distributions (TMDs) of the nucleons~\cite{Olness:tocome}.
Since similar considerations are premature for multi parton PDFs, it seems opportune to consider the 
two-current correlations in the pion ---quantities that could be related to dPDFs through moments--- with current operators
evaluated at different space points but at equal time, from a model point of view. Such correlations 
have been calculated on the lattice, {\it e.g.}, in Ref.~\cite{Burkardt:1994pw} and more recently in Refs.~\cite{Zimmermann:2017ctb,Bali:2018nde}. 
 
The novel possibility to compare results
with lattice data reinforces the theoretical relevance of model calculations of
two-current correlations in the pion.

In this paper we analyze the correlations
of two vector currents 
in the NJL model, showing
in particular their time-time component,
and attempting for this quantity a direct  comparison with lattice data.

The paper is structured as follows.
In section \ref{II} we 
define two-current correlations in the pion and their isospin decomposition, while 
in section  \ref{III} we describe the NJL evaluation scheme and we give
the explicit expressions of the results of the calculation.
In section \ref{IV} we show a comparison 
of our numerical results
with lattice data. 
Conclusions are collected in section \ref{V}.

\section{\label{II}
Definitions and isospin decomposition}

Experimental measurements of pion dPDFs appear rather challenging at 
the moment. Nevertheless, from a complementary viewpoint,
lattice evaluations of matrix elements that can be connected to
the Mellin moments of double parton distributions are possible~\cite{Diehl1}. A relevant step towards this goal has been done in Ref.~\cite{Bali:2018nde}, where lattice data
of correlation functions of two local currents in a pion, 
\begin{align}
\left\langle \pi^{k}\left(P\right)\left|\mathcal{O}_{i}^{q_{1}q_{2}}(y)\,\mathcal{O}_{j}^{q_{3}q_{4}}(0)\,\right|\pi^{k^{\prime}}\left(P\right)\right\rangle ,
\label{basic-matel1}
\end{align}
have been presented. In the expression above, the pion charges ---$k,k^{\prime}=+,-,0$---
may be different in the bra and ket state, while the four-momentum
$P$ is the same in both, and the two currents are applied at the
same time but at different spatial positions, {\it i.e.}, one has $y^{\mu}=\left(0,\,\vec{y}\right)$. %One should notice instead that 
It differs from the dPDFs which are defined 
from a similar product
of two one body operators  taken at a distance $y'^{\mu}=\left(y'^+=0,y^-\,,\vec{y'}_\perp\right)$.
Hence, the currents of interest here are 
\begin{align}
\mathcal{O}_{n}^{qq'}(y) & =\bar{q}(y)\,\Gamma_{n}\,q'(y)\,,\label{op-def}
\end{align}
where $q$ and $q'$ are $u$ or $d$ quark fields and $\Gamma_{n}$ are the Dirac bilinears $1,i\gamma_{5},\gamma^{\mu},\gamma^{\mu}\gamma_{5}$
for $n=S,P,V^{\mu},A^{\mu},$ respectively. In the present work we will
only consider the time component of the vector current, {\it i.e.}
\begin{align}
\mathcal{O}_{V^{0}}^{qq'}(y)=\bar{q}(y)\gamma^{0}q'(y)\,,\label{curr-def}
\end{align}
and we will compare our model results with Lattice
data in the case of two equal currents of this kind.

As shown in Ref. \cite{Bali:2018nde}, the matrix elements \eqref{basic-matel1}
are not all independent due to constraints from isospin symmetry and
discrete symmetries. We follow the treatment of Ref. \cite{Bali:2018nde}
defining the isosinglet current: 
\begin{align}
\mathcal{O}_{n}^{\text{s}} & =\mathcal{O}_{n}^{uu}+\mathcal{O}_{n}^{dd}\,,
\end{align}
as well as the isotriplet currents
\begin{align}
\mathcal{O}_{n}^{a} & =\bar{Q}\tau^{a}\Gamma_{n}Q\,, & a & =1,2,3\label{iso-current}
\end{align}
where $\tau^{a}$ 
are
the Pauli matrices and $Q=(u,d)$ is the 
isodoublet 
of quark fields. As a consequence, the following isospin decomposition is
obtained 
%\cite{Bali:2018nde}: 
\begin{align}
\left\langle \pi^{j}\left(P\right)\right|\mathcal{O}_{n}^{\text{s}}\left(y\right)\mathcal{O}_{n^{\prime}}^{\text{s}}\left(0\right)\left|\pi^{i}\left(P\right)\right\rangle  & =\delta^{ij}\,F_{0}(y)\,,\nonumber \\
\left\langle \pi^{j}\left(P\right)\right|\mathcal{O}_{n}^{a}\left(y\right)\mathcal{O}_{n^{\prime}}^{b}\left(0\right)\left|\pi^{i}\left(P\right)\right\rangle  & =\delta^{ab}\delta^{ij}\,F_{1}(y)+\bigl(\delta^{ai}\delta^{bj}+\delta^{aj}\delta^{bi}\bigr)\,F_{2}(y)
\nonumber \\ & +
\bigl(\delta^{ai}\delta^{bj}-\delta^{aj}\delta^{bi}\bigr)\,iF_{3}(y)\,,\nonumber \\
\left\langle \pi^{j}\left(P\right)\right|\mathcal{O}_{n}^{b}\left(y\right)\mathcal{O}_{n^{\prime}}^{\text{s}}\left(0\right)\left|\pi^{i}\left(P\right)\right\rangle  & =i\epsilon^{bij}\,G_{1}(y)\,,\nonumber \\
\left\langle \pi^{j}\left(P\right)\right|\mathcal{O}_{n}^{\text{s}}\left(y\right)\mathcal{O}_{n^{\prime}}^{b}\left(0\right)\left|\pi^{i}\left(P\right)\right\rangle  & =i\epsilon^{bij}\,G_{2}(y)\,.\label{Iso_descomp}
\end{align}
Here the real functions $F_{\ell}$ and $G_{\ell}$ ---quantities evaluated in what follows--- depend on the pion momentum $P$
and on the current indices $n,n^{\prime}$~\cite{Bali:2018nde}.

\section{\label{III}Two-current correlations in the NJL model.}

\subsection{Leading order}
We describe now the calculation of the two-current correlations,
and in particular of the functions
$F_i$, with $i=0,1,2$, and $G_l$, with $l=1,2$,
in
the NJL framework. This framework is
the most realistic model for the pseudoscalar
mesons based on a local
quantum field theory built only with quarks~\cite{Klevansky:1992qe}. 
It respects the 
realization of chiral symmetry and
gives a good description of low energy properties. Mesons are described as bound states, in
a fully covariant fashion, using the Bethe-Salpeter amplitude,
in a field theoretical framework.
In this way, Lorentz covariance is preserved.  The NJL model is a
non-renormalizable field theory and therefore a {regularization procedure has to be implemented.}
 We have performed our calculations in the Pauli--Villars  regularization
scheme ---a well established method.
The NJL model, together with its 
regularization procedure, can
be regarded as an effective theory of QCD.
Some basic features of the NJL model and details on the regularization scheme are 
reported in Appendix~\ref{App.NJL_Basic_Equations}. 
Model calculations of meson partonic structure within
this approach have a long story of successful predictions~\cite{Davidson:2001cc,Theussl:2002xp,
RuizArriola:2002bp,Courtoy:2008nf,
Courtoy:2007vy,
Courtoythesis,
Noguera:2011fv,Weigel:1999pc,
ns,
Broniowski:2017gfp,Ceccopieri:2018nop}.

Collinear parton distributions evaluated within a low energy model can be associated to a RGE scale denoted by $Q_0$. The determination of that {\it hadronic scale} reflects the degrees of freedom of the model ---here the NJL model contains only valence quarks~\cite{Stratmann:1993aw,Traini:1997jz}. The identification of such a low lying scale enables predictions from the model to measured quantities, through perturbative QCD evolution.

Let us describe the main steps of our calculation.
We define, following Eq.
\eqref{Iso_descomp},
\begin{equation}
G_{\alpha\beta}^{ij}\left(\vec{y}\right)=\left\langle \pi^{j}\left(P\right)\right|\mathcal{O}_{n}^{\beta}\left(y\right)\mathcal{O}_{n^{\prime}}^{\alpha}\left(0\right)\left|\pi^{i}\left(P\right)\right\rangle \, ,
\end{equation}
with $\alpha, \beta=\text{s},1,2,3.$ In order to proceed to the calculation
in the NJL model, we perform the Fourier transform to momentum space 
\begin{align}
G_{\alpha\beta}^{ij}\left(\vec{q}\right) & =\int d^{3}y\,e^{-i\vec{q}\cdot\vec{y}}G_{\alpha\beta}^{ij}\left(\vec{y}\right)\nonumber \\
 & =\int d^{3}y\,e^{-i\vec{q}\cdot\vec{y}}\left\langle \pi^{j}\left(P\right)\right|\left.\mathcal{O}_{n}^{\beta}\left(y\right)\mathcal{O}_{n^{\prime}}^{\alpha}\left(0\right)\right|_{y^{0}=0}\left|\pi^{i}\left(P\right)\right\rangle \, .
\label{ms}
\end{align}
We use in this expression the following definition of the state for the pion
$i$
\begin{equation}
\left|\pi^{i}\left(P\right)\right\rangle =\int d^{4}y_{1}\,d^{4}y_{2}\,\frac{d^{4}k}{\left(2\pi\right)^{4}}e^{-i\frac{1}{2}P\cdot\left(y_{1}+y_{2}\right)}\,e^{-ik\cdot\left(y_{1}-y_{2}\right)}\bar{q}\left(y_{1}\right)\phi_{\pi^{i}}\left(k,P\right)q\left(y_{2}\right)\left|0\right\rangle \, ,
\label{A.5}
\end{equation}
with $\phi_{\pi^{i}}$ the quark-pion vertex function for a $\pi^i$.
In the NJL model 
the amplitude
$\phi_{\pi^i}\left(k,P\right)$ is independent on the relative
and total quark-antiquark momenta, $k$
and $P,$ respectively, and we have 
\begin{equation}
\phi_{\pi^{i}}\left(k,P\right)=ig_{\pi qq}i\gamma_{5}\tau^{i}\,,
\label{ampli}
\end{equation}
where $g_{\pi qq}$ is the quark-pion coupling constant and $\tau^{i}$
is the isospin matrix associated to the corresponding pion $\pi^i$.

Inserting Eq.~(\ref{A.5}) in the definition~(\ref{ms}),
we obtain contributions
associated to different Feynman diagrams. The fact that $G_{\alpha\beta}^{ij}\left(\vec{y}\right)$
is defined at $y^{0}=0$ implies that 
\begin{equation}
G_{\alpha\beta}^{ij}\left(\vec{q}\right)=\int\frac{dq^{0}}{2\pi}\,\mathcal{T}\label{A.9-1} \, ,
\end{equation}
where $\mathcal{T}$ is the Feynman amplitude corresponding to the diagram. 

\begin{figure}
\begin{centering}
\includegraphics[scale=0.3]{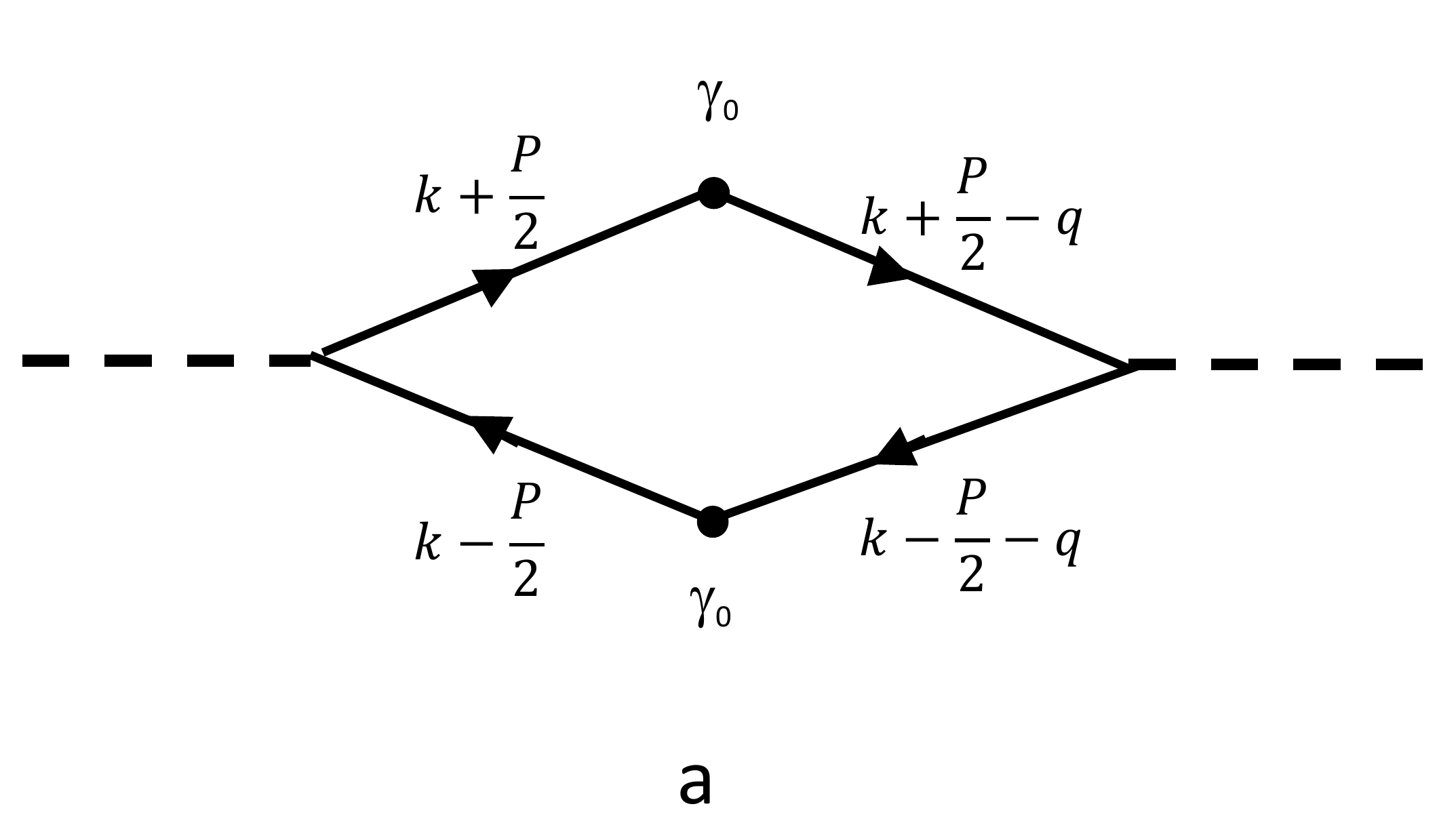}
\includegraphics[scale=0.4]{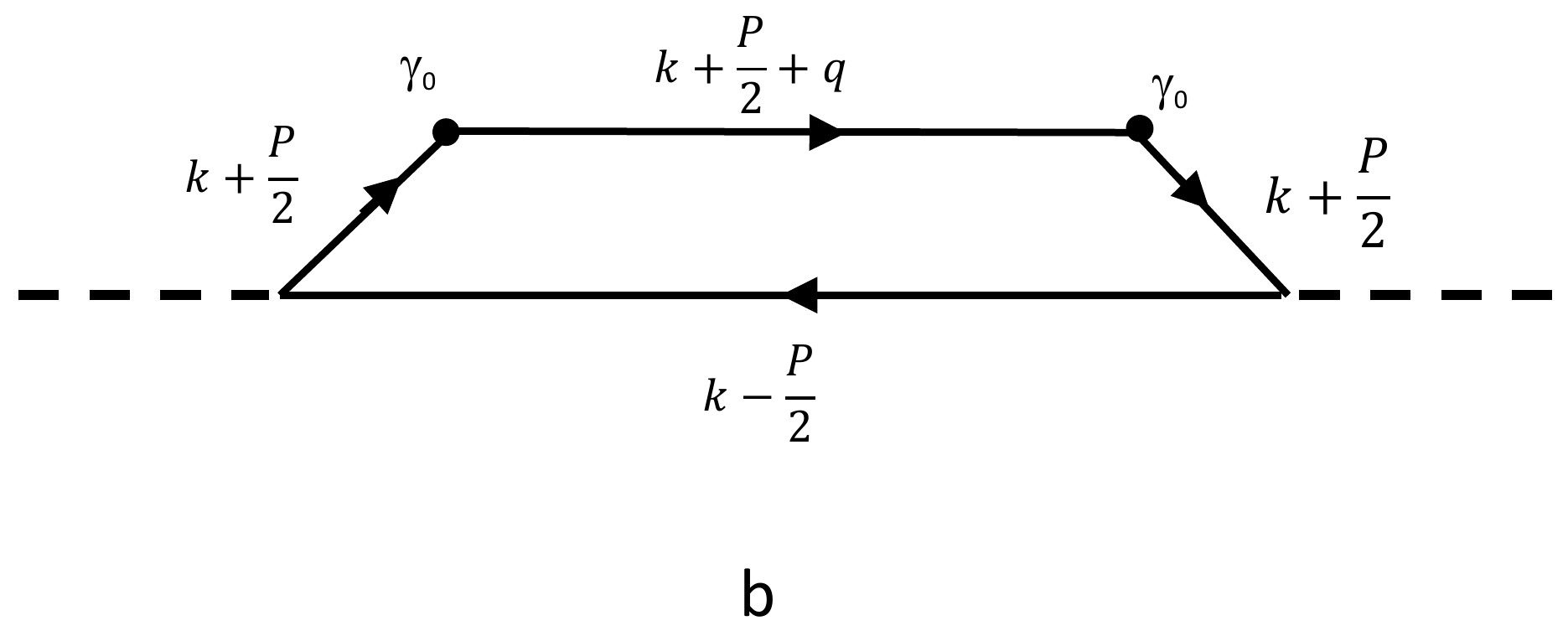}
\par\end{centering}
\caption{Examples of the two types of diagrams contributing to the leading
order of the time component of the two local vector current correlations: {\it rhombus} (a) and {\it trapezoid} (b).}
\label{Fig Leading local diagrams}
\end{figure}

The two contributions to be considered at leading order are shown in Fig.
\ref{Fig Leading local diagrams}. In the following, we will call them
{\it rhombus} and {\it trapezoid} diagrams,
respectively. Therefore one has 
\begin{eqnarray}
G_{\alpha\beta}^{\left(0\right)ij}(\vec{q})=G_{\alpha\beta}^{\left(r\right)ij}(\vec{q})+G_{\alpha\beta}^{\left(t\right)ij}(\vec{q}) \, ,
\end{eqnarray}
where $G_{\alpha\beta}^{\left(r\right)ij}(\vec{q})$ and $G_{\alpha\beta}^{\left(t\right)ij}(\vec{q})$
stand for contributions from the rhombus and trapezoid diagrams, respectively,
which read

\begin{eqnarray}
G_{\alpha\beta}^{\left(r\right)ij}(\vec{q}) & = & \int\frac{d^{4}k}{(2\pi)^{4}}\frac{dq^{0}}{2\pi}(-)\left\{ \text{Tr}\left[\bar{\phi}_{\pi^{j}}iS_{F}\left(\frac{P}{2}+k+q\right)\tau^{\alpha}\Gamma_{n^{\prime}}iS_{F}\left(k+\frac{P}{2}\right)\right.\right.\nonumber \\
& \times & \left.\left.\phi_{\pi^{i}}iS_{F}\left(k-\frac{P}{2}\right)\tau^{\beta}\Gamma_{n}iS_{F}\left(k+q-\frac{P}{2}\right)\right] \right .
\nonumber \\ 
& + & \left . (q^{\mu}\leftrightarrow-q^{\mu},\left(\alpha,n^{\prime}\right)\leftrightarrow\left(\beta,n\right))\right\} \label{romboG} \, ,
\end{eqnarray}

\begin{eqnarray}
G_{\alpha\beta}^{\left(t\right)ij}(\vec{q}) & = & \int\frac{d^{4}k}{(2\pi)^{4}}\frac{dq^{0}}{2\pi}(-)\left\{ \text{Tr}\left[\bar{\phi}_{\pi^{j}}iS_{F}\left(\frac{P}{2}+k\right)\tau^{\beta}\Gamma_{n}iS_{F}\left(k+\frac{P}{2}+q\right)\right.\right.\nonumber \\
 & \times & \left.\left.\tau^{\alpha}\Gamma_{n^{\prime}}iS_{F}\left(k+\frac{P}{2}\right)\phi_{\pi^{i}}iS_{F}\left(k-\frac{P}{2}\right)\right]\right.\nonumber \\
 & + & \left.\text{Tr}\left[\bar{\phi}_{\pi^{j}}iS_{F}\left(\frac{P}{2}+k\right)\phi_{\pi^{i}}\,iS_{F}\left(k-\frac{P}{2}\right)\tau^{\beta}\Gamma_{n}\right.\right.\nonumber \\
 & \times & \left.\left.iS_{F}\left(k-\frac{P}{2}+q\right)\tau^{\alpha}\Gamma_{n^{\prime}}iS_{F}\left(k-\frac{P}{2}\right)\right] \right.\nonumber \\
& + & \left. (q^{\mu}\leftrightarrow-q^{\mu},\left(\alpha,m^{\prime}\right)\leftrightarrow\left(\beta,m\right)) \right\}.
 \label{trapecioG}
\end{eqnarray}
Here $\tau^{\alpha}=\left(\tau^{\text{s}}=1,\,\vec{\tau}\right),$
with $\vec{\tau}$ the standard Pauli matrices. 

According to their isospin structure, which is determined by $\phi_{i}$ and
$\tau^{\alpha}$ terms, the two expressions above yield four types
of contributions once the isospin traces are evaluated ---a straightforward lengthy procedure. For the {\it rhombus} contribution, they read

\begin{eqnarray}
G_{ss}^{\left(r\right)ij}(\vec{q}) & = & \delta_{ij}2N_{c}(ig_{\pi qq})^{2}[I_{\Gamma}^{r}(m,\vec{q})+I_{\Gamma}^{r}(m,-\vec{q})\,,\nonumber \\
G_{sa}^{\left(r\right)ij}(\vec{q}) & = & 2i\epsilon_{jia}N_{c}(ig_{\pi qq})^{2}[I_{\Gamma}^{r}(m,\vec{q})-I_{\Gamma}^{r}(m,-\vec{q})]\,,\nonumber \\
G_{as}^{\left(r\right)ij}(\vec{q}) & = & 2i\epsilon_{ija}N_{c}(ig_{\pi qq})^{2}[I_{\Gamma}^{r}(m,\vec{q})-I_{\Gamma}^{r}(m,-\vec{q})]\,,\nonumber \\
G_{ab}^{\left(r\right)ij}(\vec{q}) & = & (\delta_{ja}\delta_{ib}-\delta_{ji}\delta_{ab}+\delta_{jb}\delta_{ia})2N_{c}(ig_{\pi qq})^{2}[I_{\Gamma}^{r}(m,\vec{q})+I_{\Gamma}^{r}(m,-\vec{q})]\,,\label{rombo4}
\end{eqnarray}
with

\begin{eqnarray}
I_{\Gamma}^{r}(m,\vec{q}) & = & \int\frac{d^{4}k}{(2\pi)^{4}}
\frac{dq_{0}}{2\pi}\text{tr}\left[ 
\gamma_{5} 
S_{F}\left(k+q+\frac{P}{2}\right)\Gamma S_{F}\left(k+\frac{P}{2}\right)
\right .
\nonumber \\ 
& \times & 
\left .
\gamma_{5}  
S_{F}\left(k-\frac{P}{2}\right)\Gamma S_{F}\left(k+q-\frac{P}{2}\right)\right]\,.
\end{eqnarray}
In the equations above, $m$ is the quark mass and we have assumed that $\Gamma_{n}=\Gamma_{n^{\prime}}=\Gamma.$
For the {\it trapezoid} contribution, they read
\begin{eqnarray}
G_{ss}^{\left(t\right)ij}(\vec{q}) & = & \delta_{ij}2N_{c}(ig_{\pi qq})^{2}[I_{\Gamma}^{t_{1}}(m,\vec{q})+I_{\Gamma}^{t_{1}}(m,-\vec{q})+I_{\Gamma}^{t_{2}}(m,\vec{q})+I_{\Gamma}^{t_{2}}(m,-\vec{q})]\,,\nonumber \\
G_{sa}^{\left(t\right)ij}(\vec{q}) & = & 2i\epsilon_{jia}N_{c}(ig_{\pi qq})^{2}[-I_{\Gamma}^{t_{1}}(m,\vec{q})+I_{\Gamma}^{t_{1}}(m,-\vec{q})+I_{\Gamma}^{t_{2}}(m,\vec{q})-I_{\Gamma}^{t_{2}}(m,-\vec{q})]\,,\nonumber \\
G_{as}^{\left(t\right)ij}(\vec{q}) & = & 2i\epsilon_{jia}N_{c}(ig_{\pi qq})^{2}[-I_{\Gamma}^{t_{1}}(m,\vec{q})+I_{\Gamma}^{t_{1}}(m,-\vec{q})+I_{\Gamma}^{t_{2}}(m,\vec{q})-I_{\Gamma}^{t_{2}}(m,-\vec{q})]\,,\nonumber \\
G_{ab}^{\left(t\right)ij}(\vec{q}) & = & 2N_{c}(ig_{\pi qq})^{2}\{\delta_{ji}\delta_{ab}[I_{\Gamma}^{t_{1}}(m,\vec{q})+I_{\Gamma}^{t_{1}}(m,-\vec{q})+I_{\Gamma}^{t_{2}}(m,\vec{q})+I_{\Gamma}^{t_{2}}(m,-\vec{q})]\label{trapecio4}\\
 & + & (\delta_{jb}\delta_{ai}-\delta_{ja}\delta_{bi})[I_{\Gamma}^{t_{1}}(m,\vec{q})-I_{\Gamma}^{t_{1}}(m,-\vec{q})-I_{\Gamma}^{t_{2}}(m,\vec{q})+I_{\Gamma}^{t_{2}}(m,-\vec{q})]\}\,,\nonumber 
\end{eqnarray}
with

\begin{eqnarray}
I_{\Gamma}^{t_{1}}(m,\vec{q})& = & \int\frac{d^{4}k}{(2\pi)^{4}}\frac{dq^{0}}{2\pi}
\, \text{tr}\left[\gamma_{5}S_{F}\left(k+\frac{P}{2}\right)
\Gamma S_{F}\left(k+\frac{P}{2}+q\right)
\right .
\nonumber \\
& \times & \left .
\Gamma S_{F}\left(k+\frac{P}{2}\right)\gamma_{5}S_{F}\left(k-\frac{P}{2}\right)\right]\,,
\end{eqnarray}

and

\begin{eqnarray}
I_{\Gamma}^{t_{2}}(m,\vec{q}) & = & \int\frac{d^{4}k}{(2\pi)^{4}}\frac{dq^{0}}{2\pi}\text{tr}\left[\gamma_{5}S_{F}\left(k+\frac{P}{2}\right)\gamma_{5}S_{F}\left(k-\frac{P}{2}\right)
\right .
\nonumber \\
& \times & \left .
\Gamma S_{F}\left(k-\frac{P}{2}+q\right)\Gamma S_{F}\left(k-\frac{P}{2}\right)\right]\,.
\end{eqnarray}

By comparing Eq. (\ref{rombo4}) and (\ref{trapecio4}) with Eq. (\ref{Iso_descomp})
we obtain the relations,

\begin{eqnarray}
F_{0}(\vec{q}) & = & X_{\Gamma}^{r}(\vec{q})+X_{\Gamma}^{t}(\vec{q})\,,\nonumber \\
F_{1}(\vec{q}) & = & -X_{\Gamma}^{r}(\vec{q})+X_{\Gamma}^{t}(\vec{q})\,,\nonumber \\
F_{2}(\vec{q}) & = & X_{\Gamma}^{r}(\vec{q})\,,\nonumber \\
iF_{3}(\vec{q}) & = & -Z_{\Gamma}^{t}(\vec{q})\,,\nonumber \\
G_{1}(\vec{q}) & = & -Y_{\Gamma}^{r}(\vec{q})-Y_{\Gamma}^{t}(\vec{q})\,,\nonumber \\
G_{2}(\vec{q}) & = & Y_{\Gamma}^{r}(\vec{q})-Y_{\Gamma}^{t}(\vec{q})\,,\label{ottorel}
\end{eqnarray}
with 
\begin{eqnarray}
X_{\Gamma}^{r}(\vec{q}) & = & 2N_{c}(ig_{\pi qq})^{2}[I_{\Gamma}^{r}(m,\vec{q})+I_{\Gamma}^{r}(m,-\vec{q})]\,,\nonumber \\
Y_{\Gamma}^{r}(\vec{q}) & = & 2N_{c}(ig_{\pi qq})^{2}[I_{\Gamma}^{r}(m,\vec{q})-I_{\Gamma}^{r}(m,-\vec{q})]\,,\nonumber \\
X_{\Gamma}^{t}(\vec{q}) & = & 2N_{c}(ig_{\pi qq})^{2}[I_{\Gamma}^{t_{1}}(m,\vec{q})+I_{\Gamma}^{t_{2}}(m,\vec{q})+I_{\Gamma}^{t_{1}}(m,-\vec{q})+I_{\Gamma}^{t_{2}}(m,-\vec{q})]\,,\nonumber \\
Y_{\Gamma}^{t}(\vec{q}) & = & 2N_{c}(ig_{\pi qq})^{2}[-I_{\Gamma}^{t_{1}}(m,\vec{q})+I_{\Gamma}^{t_{2}}(m,\vec{q})+I_{\Gamma}^{t_{1}}(m,-\vec{q})-I_{\Gamma}^{t_{2}}(m,-\vec{q})]\,,\nonumber \\
Z_{\Gamma}^{t}(\vec{q}) & = & 2N_{c}(ig_{\pi qq})^{2}[I_{\Gamma}^{t_{1}}(m,\vec{q})-I_{\Gamma}^{t_{2}}(m,\vec{q})-I_{\Gamma}^{t_{1}}(m,-\vec{q})+I_{\Gamma}^{t_{2}}(m,-\vec{q})]\,.\label{xyz}
\end{eqnarray}
In Ref. \cite{Bali:2018nde} it is shown that $iF_{3}\left(\vec{q}\right)$
is different from zero only if $\Gamma_{n}\neq\Gamma_{n^{\prime}},$
and charge conjugation constraints imply that $G_{\ell}\left(\vec{q}\right)=0$
for the cases $\Gamma_{n}=\Gamma_{n^{\prime}}.$ We observe that these
results are automatically reproduced if $I_{\Gamma}^{r}(m,\vec{q})$
and $I_{\Gamma}^{t_{i}}(m,\vec{q})$ are functions of $\left|\vec{q}\right|$
only.
We will see later on that this fact actually occurs
in our calculation.

The integrals over $q^{0}$ and $k^{0}$ in
the expressions
$I_{\Gamma}^{r,t_{i}}(\vec{q})$
can be easily evaluated using the poles of the propagators. For instance,
when $\Gamma=V^{0}=\gamma^{0}$ we obtain
\begin{equation}
I_{V^{0}}^{r}(m,\vec{q})=  \bigintss\frac{d^{3}k}{(2\pi)^{3}}\frac{4\left[E_{\left|\vec{k}\right|}^{2}\left(4\,E_{\left|\vec{k}+\vec{q}\right|}^{2}+m_{\pi}^{2}\right)+m_{\pi}^{2}\,\vec{k}\cdot\vec{q}\right]}{E_{\left|\vec{k}\right|}E_{\left|\vec{k}+\vec{q}\right|}\left(4\,E_{\left|\vec{k}\right|}^{2}-m_{\pi}^{2}\right)\left(4\,E_{\left|\vec{k}+\vec{q}\right|}^{2}-m_{\pi}^{2}\right)} \, ,
\end{equation}
where $E_{v}=\sqrt{v^{2}+m^{2}}.$ Now, the angular integration can be
performed, arriving to

\begin{eqnarray}
I_{V^{0}}^{r}(m,q) & = & \bigintsss\frac{dk}{(2\pi)^{2}}\frac{k}{qE_{k}}\left\{ (E_{ k+ q }-E_{k - q})\left(1+\frac{3}{2}\frac{m_{\pi}^{2}}{
4E_{k}^{2}-m_{\pi}^{2}}\right)\right.\nonumber \\
 & + & \left.m_{\pi}\frac{32E_{k}^{2}-2E_{k+ q}^{2}-
 2E_{k-q}^{2}+m_{\pi}^{2}}{8(4E_{ k}^{2}-m_{\pi}^{2})}\ln\frac{(2E_{ k-  q}+m_{\pi})
 (2E_{k+ q}-m_{\pi})}{(2E_{ k-q}-m_{\pi})(2E_{k+ q}+m_{\pi})}\right\} \,,
\end{eqnarray}
where $k=|\vec{k}|$ and $q=\left|\vec{q}\right|.$ In
a similar way we obtain

\begin{eqnarray}
I_{V^{0}}^{t}(m,q) & = & I_{V^{0}}^{t_{1}}(m,q)+I_{V^{0}}^{t_{2}}(m,q)=\nonumber \\
 & = & \int\frac{dk}{(2\pi)^{2}}\left(-\frac{k}{qE_{k}}\right)(E_{k+q}-E_{k-q})\left\{ 1-\frac{1}{3}\frac{(E_{k+q}-E_{k-q})^{2}}{4E_{k}^{2}-m_{\pi}^{2}}\right.\nonumber \\
 & - & \left.\frac{m_{\pi}^{2}}{3(4E_{k}^{2}-m_{\pi}^{2})^{2}}\left[36E_{k}^{2}-2(E_{k+q}-E_{k-q})^{2}-3m_{\pi}^{2}\right]\right\} \,.
\end{eqnarray}

After regularization, the same integrals become

\begin{eqnarray}
\bar{I}_{V^{0}}^{r}(q)=\sum_{i=0}^{2}c_{i}I_{V^{0}}^{r}(M_{i},q)\,,\label{ivb}
\end{eqnarray}

\begin{eqnarray}
\bar{I}_{V^{0}}^{t}(q)=\sum_{i=0}^{2}c_{i}I_{V^{0}}^{t}(M_{i},q)\,.\label{itb}
\end{eqnarray}
One can see that these integrals do not depend on the direction of
$\vec{q}$, as anticipated, and therefore Eqs. \eqref{xyz} become 

\begin{eqnarray}
X_{V^{0}}^{r}(q) & = & 4N_{c}(ig_{\pi qq})^{2}[\bar{I}_{V^{0}}^{r}(q)]\,,\nonumber \\
X_{V^{0}}^{t}(q) & = & 4N_{c}(ig_{\pi qq})^{2}[\bar{I}_{V^{0}}^{t}(q)]\,,\nonumber \\
Y_{V^{0}}^{r}(q) & = & Y_{V^{0}}^{t}(q)=Z_{V^{0}}^{t}(q)=0\,.\label{newxyzV0}
\end{eqnarray}

We conclude this section observing that
the above equations define the leading order contributions
to the functions \eqref{ottorel} in the $\Gamma=\gamma^0=V^0$ case:

\begin{eqnarray}
F_{0}({q}) & = & X_{V^0}^{r}({q})+X_{V^0}^{t}({q})\,,\nonumber \\
F_{1}({q}) & = & -X_{V^0}^{r}({q})+X_{V^0}^{t}({q})\,,\nonumber \\
F_{2}(\vec{q}) & = & X_{V^0}^{r}({q})\,,\nonumber \\
iF_{3}({q}) & = & 0 \,,\nonumber \\
G_{1}({q}) & = & 0 \,,\nonumber \\
G_{2}({q}) & = & 0 \,.
\label{summary}
\end{eqnarray}

\subsection{Dressing the vertex and other contributions}

\begin{figure}
\begin{centering}
\includegraphics[scale=0.5]{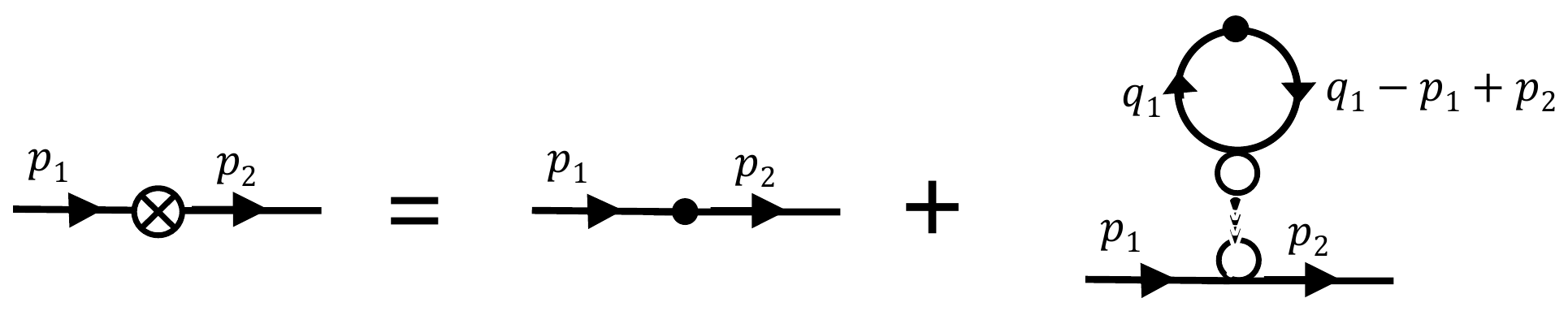}
\par\end{centering}
\caption{Dressed local vertex.}
\label{Fig_dressed_local_vertex}
\end{figure}

In addition to the contribution
we have just discussed, one has to consider that,
in the present NJL scheme,
the local vertex can be dressed by the chiral interaction through
the diagrams depicted in Fig. \ref{Fig_dressed_local_vertex}. This
dressing, which could produce corrections to the results obtained in the previous subsection, 
implies to change the bare vertex $\tau^{\alpha}\Gamma$ as follows, for the  isoscalar and isovector vertices respectively
\begin{align}
\Gamma & \rightarrow\Gamma+\frac{2ig}{1-2g\,\Pi_{S}\left(q^{2}\right)}\left(-2N_{c}\right)\int\frac{d^{4}q_{1}}{\left(2\pi\right)^{4}}\text{tr}\left[iS_{F}\left(q_{1}\right)\,iS_{F}\left(q_{1}-q\right)\,\Gamma\right] \, ,\nonumber \\
\label{Dres_vert_lineal}\\
\tau^{a}\Gamma & \rightarrow\tau^{a}\Gamma+i\gamma_{5}\tau^{a}\frac{2ig}{1-2g\,\Pi_{PS}\left(q^{2}\right)}\left(-2N_{c}\right)\int\frac{d^{4}q_{1}}{\left(2\pi\right)^{4}}\text{tr}\left[iS_{F}\left(q_{1}\right)\,i\gamma_{5}\,iS_{F}\left(q_{1}-q\right)\,\Gamma\right]\, , \nonumber 
\end{align}
where $q^{\mu}=p_{1}^{\mu}-p_{2}^{\mu}=\left(0,\vec{q}\right).$

In the vector current case, only the isoscalar current can
be dressed in principle. However, it is also easily proved that such a contribution vanishes, as
\begin{equation}
\int\frac{d^{4}q_{1}}{\left(2\pi\right)^{4}}\text{tr}\left[iS_{F}\left(q_{1}\right)\,iS_{F}\left(q_{1}-q\right)\,\gamma^{0}\right]=0\,.
\end{equation}
Therefore, in the particular case of the vector current under investigation here, there is no dressing of the vertex.

\begin{figure}
\centering{}
\includegraphics[scale=0.5]{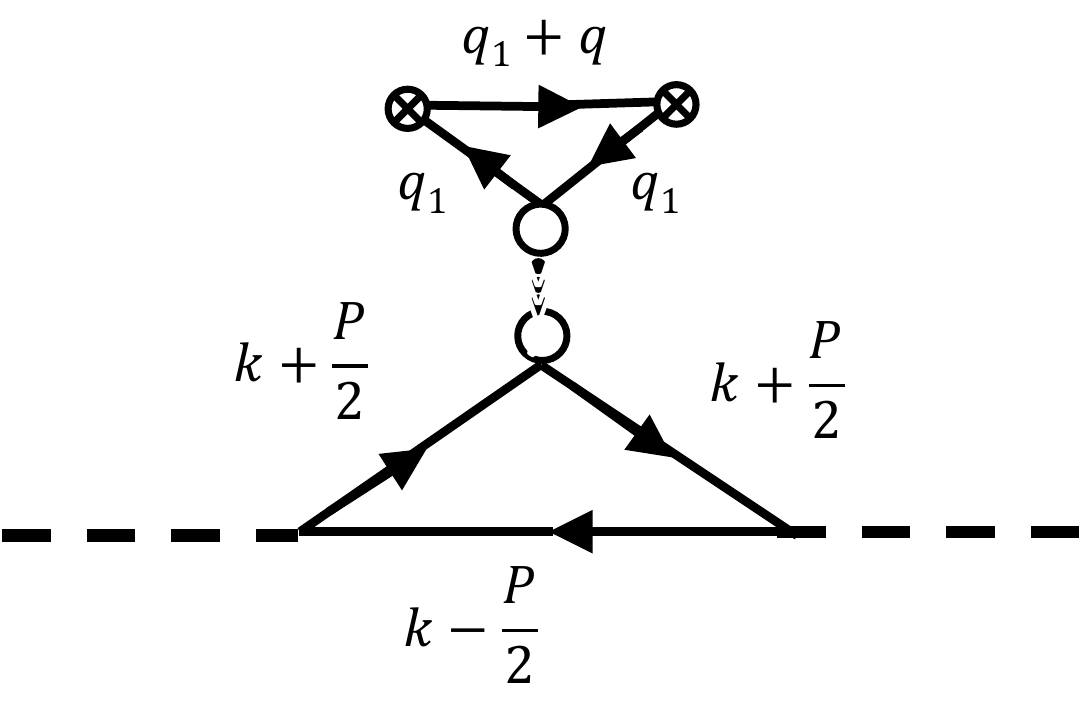}
\caption{Other type of contribution originated by the dressing of the currents. }
\label{Fig_p_diagram}
\end{figure}

Finally, we consider the diagram depicted in Fig.~\ref{Fig_p_diagram}. 
This contribution can be written in the form ---$(p)$ stands for {\it pole contribution}, 
\begin{equation}
G_{\alpha\beta}^{\left(p\right)ij}\left(\vec{q}\right)=\mathbb{M}_{ij}\,\frac{2ig}{1-2g\,\Pi_{S}\left(0\right)}\,\mathbb{N_{\alpha\beta}} \, ,
\end{equation}
with
\begin{align}
\mathbb{M}_{ij}= & -\int\frac{d^{4}\ell}{\left(2\pi\right)^{4}}\text{Tr}\left[\bar{\phi}_{\pi^{j}}\,iS_{F}\left(P+\ell\right)\,iS_{F}\left(P+\ell\right)\,\phi_{\pi^{i}}\,iS_{F}\left(\ell\right)\right]\nonumber \\
 & -\int\frac{d^{4}\ell}{\left(2\pi\right)^{4}}\text{Tr}\left[\bar{\phi}_{\pi^{j}}\,iS_{F}\left(\ell\right)\,\phi_{\pi^{i}}\,iS_{F}\left(\ell-P\right)\,iS_{F}\left(\ell-P\right)\right]
 \, ,
\end{align}
and
\begin{equation}
\mathbb{N}_{\alpha\beta}=-\int\frac{d^{4}k}{\left(2\pi\right)^{4}}\text{Tr}\left[iS_{F}\left(k\right)\,iS_{F}\left(k\right)\,\tau^{\beta}\Gamma\,iS_{F}\left(k+q\right)\,\tau^{\alpha}\Gamma\right]+\left(q^{\mu}\leftrightarrow-q^{\mu},\alpha\leftrightarrow\beta\right) \, .
\end{equation}
The direct calculation gives, for $\mathbb{M}_{ij}$,
\begin{equation}
\mathbb{M}_{ij}=\delta_{ij}\,16N_{c}g_{\pi qq}^{2}\,m\,\left[m_{\pi}^{2}\,I_{3}\left(m_{\pi}^{2}\right)-I_{2}\left(0\right)\right] \, ,
\end{equation}
with $I_{2}\left(q^{2}\right)$ and $I_{3}\left(q^{2}\right)$ defined
in App. \ref{App.NJL_Basic_Equations}. The two quark intermediate state
is described by the quantity, 
\begin{equation}
\frac{2ig}{1-2g\,\Pi_{S}\left(0\right)}=\frac{-i}{4N_{c}\left[m_{\pi}^{2}I_{2}\left(m_{\pi}^{2}\right)+4m^{2}I_{2}\left(0\right)\right]} \, ,
\end{equation}
and for $\mathbb{N}_{\alpha\beta}$ we have
\begin{equation}
\mathbb{N}_{\alpha\beta}=\delta_{\alpha\beta}\,i\,2N_{c}\left[I_{\Gamma}^{p}\left(m,\vec{q}\right)+I_{\Gamma}^{p}\left(m,-\vec{q}\right)\right] \, ,
\end{equation}
with
\begin{equation}
I_{\Gamma}^{p}\left(m,\vec{q}\right)=\int\frac{d^{4}k}{\left(2\pi\right)^{4}}\text{tr}\left[S_{F}\left(k\right)\,S_{F}\left(k\right)\,\Gamma\,S_{F}\left(k+q\right)\,\Gamma\right] \, .
\end{equation}
Putting all these results together, we obtain
\begin{equation}
G_{\alpha\beta}^{\left(p\right)ij}\left(\vec{q}\right)=
\delta_{ij}\delta_{\alpha\beta}\,2N_{c}\, (ig_{\pi qq})^2\,(- 4m^2)\,\frac{\left[m_{\pi}^{2}\,I_{3}\left(m_{\pi}^{2}\right)-I_{2}\left(0\right)\right]}{\left[m_{\pi}^{2}I_{2}\left(m_{\pi}^{2}\right)+4m^{2}I_{2}\left(0\right)\right]}\left[I_{\Gamma}^{p}\left(m,\vec{q}\right)+I_{\Gamma}^{p}\left(m,-\vec{q}\right)\right]
\, .
\label{polo4}
\end{equation}
By comparing to the structures for the isospin amplitudes defined in  Eqs.~(\ref{Iso_descomp}), it is observed that the {\it pole} contributions add to the bare leading-order results in Eqs.~(\ref{ottorel}) in the following manner
\begin{eqnarray}
F_{0}(\vec{q}) & = & X_{\Gamma}^{r}(\vec{q})+X_{\Gamma}^{t}(\vec{q})+X_{\Gamma}^{p}(\vec{q})\,,\nonumber \\
F_{1}(\vec{q}) & = & -X_{\Gamma}^{r}(\vec{q})+X_{\Gamma}^{t}(\vec{q})+X_{\Gamma}^{p}(\vec{q})\,,
\nonumber \\
F_{2}(\vec{q}) & = & X_{\Gamma}^{r}(\vec{q})\,,
\label{ottorel1}
\end{eqnarray}
with 

\begin{equation}
X_{\Gamma}^{p}(\vec{q})=8N_{c}g_{\pi qq}^{2}\,m^2\,\frac{\left[m_{\pi}^{2}\,I_{3}\left(m_{\pi}^{2}\right)-I_{2}\left(0\right)\right]}{\left[m_{\pi}^{2}I_{2}\left(m_{\pi}^{2}\right)+4m^{2}I_{2}\left(0\right)\right]}\left[I_{\Gamma}^{p}\left(m,\vec{q}\right)+I_{\Gamma}^{p}\left(m,-\vec{q}\right)\right] \, .
\end{equation}

In the case $\Gamma=\gamma^{0},$
shown in the following as an example, $I_{\Gamma}^{p}\left(m,\vec{q}\right)$
is a function of $q=\left|\vec{q}\right|$,
\begin{align}
I_{V^{0}}^{p}\left(m,q\right) & %=I_{red}^{p}\left(m,q\right) \, ,\\
%I_{red}^{p}\left(m,q\right) & 
=\int\frac{dk}{\left(2\pi\right)^{2}}\frac{-k}{12\,q\,E_{k}^{3}}\left(E_{k+q}-E_{k-q}\right)^{3} \, , \nonumber 
\end{align}
and, after regularization, the last integral becomes
\begin{equation}
\bar{I}_{V^{0}}^{\,p}\left(q\right)=\sum_{i=0}^{2}c_{i}\,I_{V^{0}}^{p}\left(M_{i},q\right) \, .
\end{equation}
Putting all these results together we have finally

\begin{equation}
\label{xvp}
X_{V^{0}}^{p}\left(q\right)=16N_{c}g_{\pi qq}^{2}\,m^{2}\,\frac{\left[m_{\pi}^{2}\,I_{3}\left(m_{\pi}^{2}\right)-I_{2}\left(0\right)\right]}{\left[m_{\pi}^{2}I_{2}\left(m_{\pi}^{2}\right)+4m^{2}I_{2}\left(0\right)\right]}\bar{I}_{V^{0}}^{\,p}\left(q\right) \, ,
\end{equation}
and the final result for the functions
under investigation, including all the contributions evaluated so far, can be 
summarized as follows
\begin{eqnarray}
F_{0}({q}) & = & X_{V^0}^{r}({q})+X_{V^0}^{t}({q})+
X_{V^0}^{p}({q})\,,\nonumber \\
F_{1}({q}) & = & -X_{V^0}^{r}({q})+X_{V^0}^{t}({q})+X_{V^0}^{p}({q})\,,
\nonumber \\
F_{2}({q}) & = & X_{V^0}^{r}({q})\,.
\label{summ2}
\end{eqnarray}

\begin{figure}
\centering{}
\includegraphics[scale=0.6]{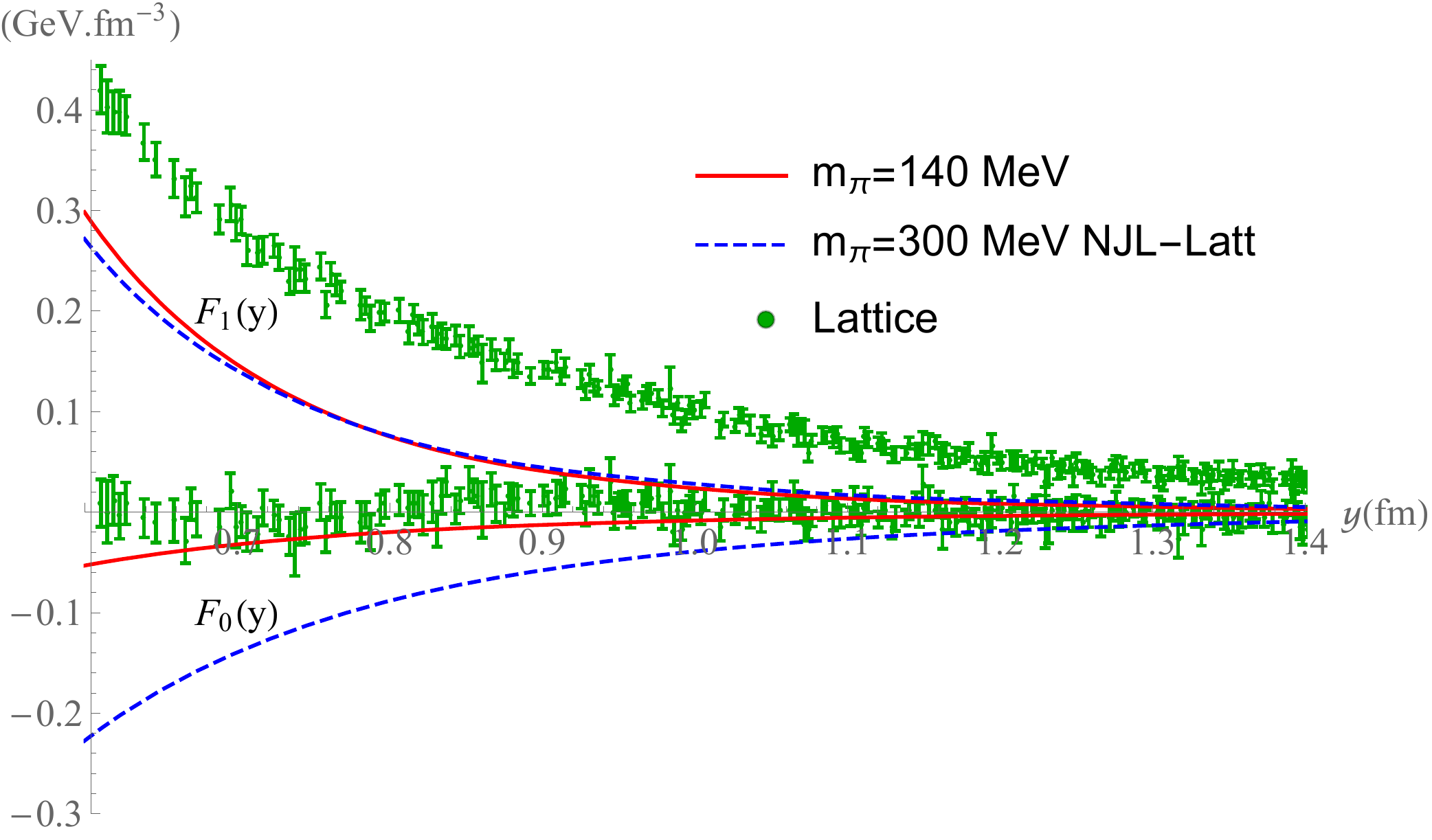} 
\caption{The quantities $F_0$ and $F_1$, Eqs.~\eqref{summ2},
in coordinate space, for various choices of the NJL parameters ---given in App.~\ref{App.NJL_PV_regularization}--- corresponding to a different
pion mass, to allow a proper comparison with lattice data~\cite{Bali:2018nde}
(green points with error band), for $\Gamma=\gamma^0$.}
\label{Fig:7} 
\end{figure}

\begin{figure}
\centering{}\includegraphics[scale=0.65]{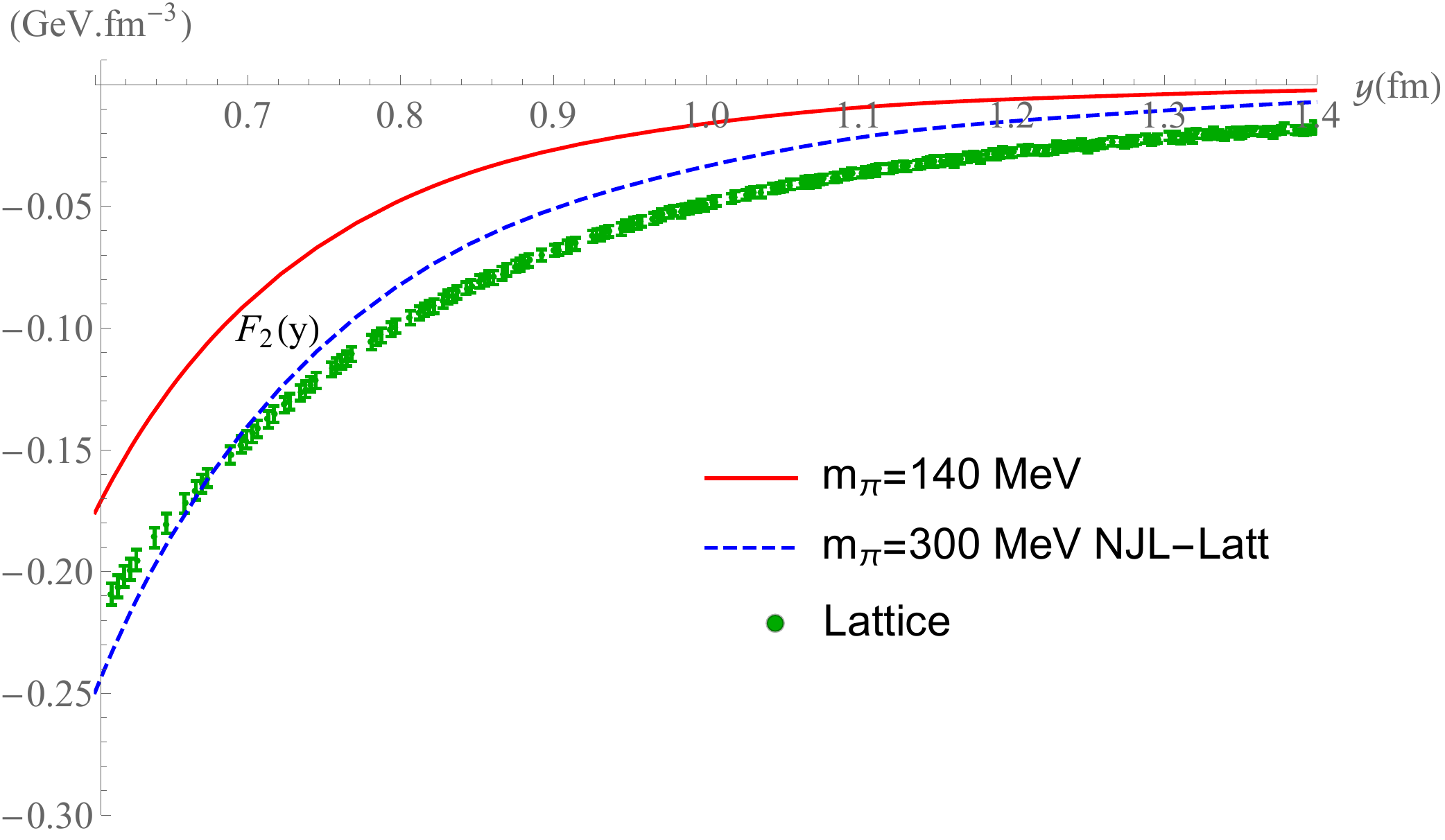} \caption{
Same as in Fig. 4, but for the function
$F_2$ in Eqs. \eqref{summ2}.
}
\label{Fig:10} 
\end{figure}

\begin{figure}[htbp]
\centering{}\includegraphics[scale=0.7]{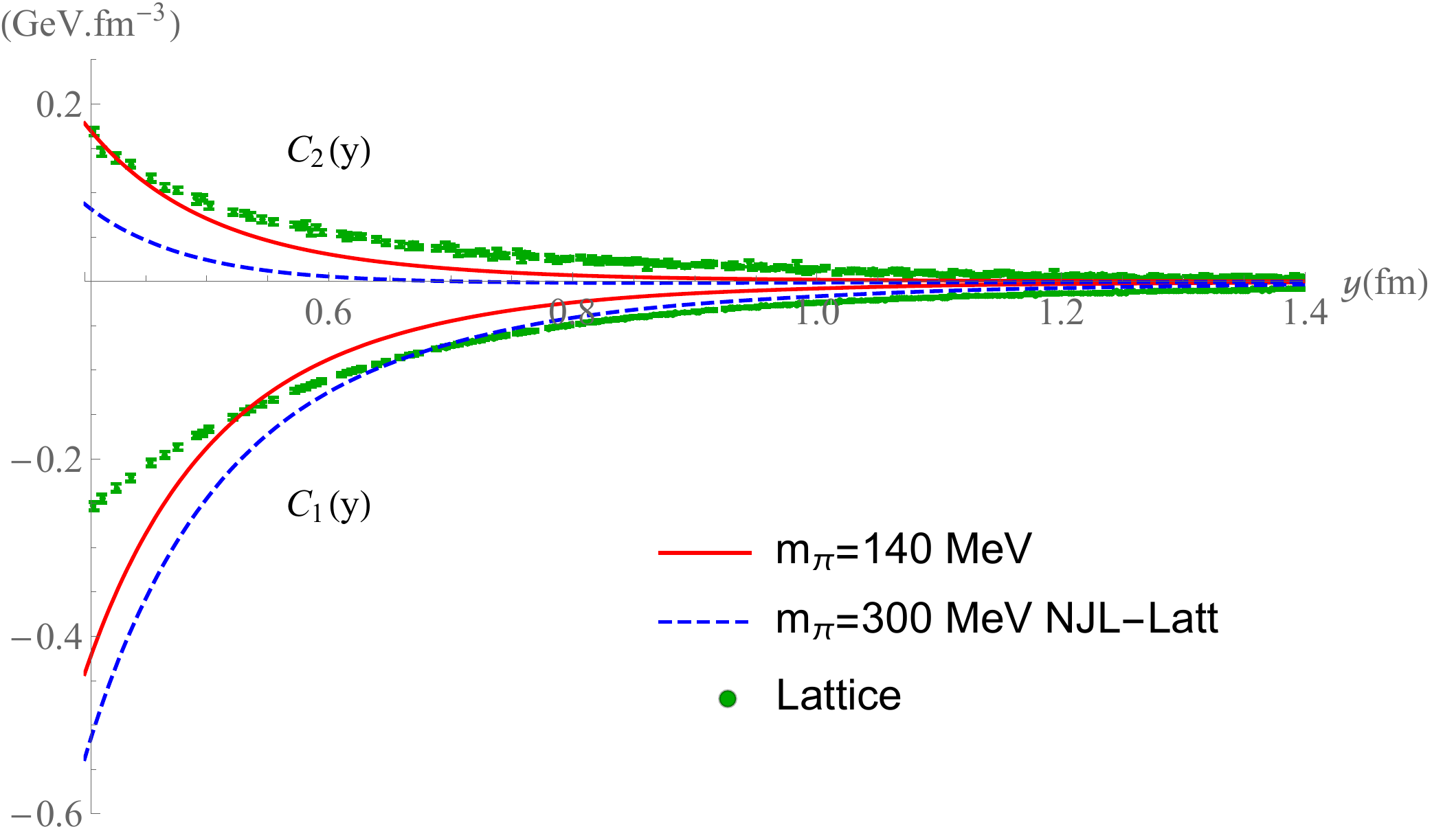} \caption{The functions $C_{1}$ and $C_{2}$, Eq. 
\eqref{duerel}, for various choices of the NJL parameters ---given in App.~\ref{App.NJL_PV_regularization}---  corresponding to a different
pion mass to allow a proper comparison with lattice data
\cite{Bali:2018nde}
(green points with error band), for $\Gamma=\gamma^0$, in coordinate space.}
\label{Fig:8} 
\end{figure}

\begin{figure}
\centering{}\includegraphics[scale=0.8]{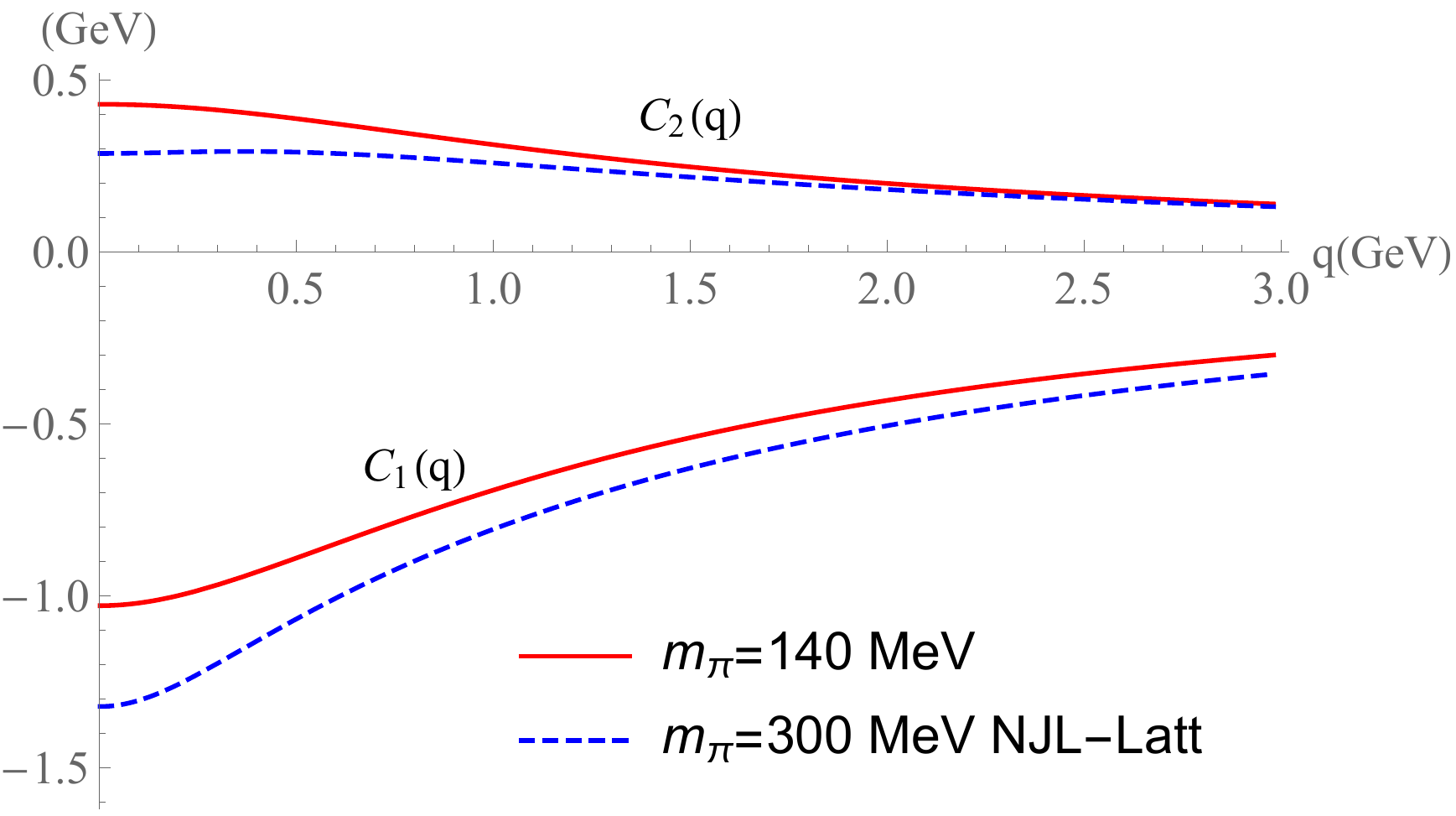} \caption{Same as in Fig.~\ref{Fig:8}, but in
momentum space.}
\label{Fig:9} 
\end{figure}

\section{\label{IV} Comparison with lattice data}

The comparison of our results with the lattice evaluation of Ref.~\cite{Bali:2018nde}
in the vector-vector case
is presented in Figs.~\ref{Fig:7}-\ref{Fig:9}.

The quantities $F_0$ and $F_1$ in Eqs. \eqref{summ2} ---evaluated using Eqs. 
\eqref{newxyzV0} and \eqref{xvp}---
are shown for $\Gamma=\gamma^0$ in Fig. \ref{Fig:7}, in coordinate space, for various choices of the parameters
of the NJL model corresponding to a different 
pion mass. The sets of model parameters corresponding to different values of the pion mass are reported in Appendix \ref{App.NJL_Basic_Equations}. The change of the full parameter set allows a proper comparison with lattice data.
The blue dashed curves ---labelled
{\it $m_\pi=$~300~MeV~NJL--Latt}---
have to be compared with the lattice results ---in green. Those lattice data correspond to  Figs. 20 (c)
and (a) of Ref.~\cite{Bali:2018nde}, respectively. 
A qualitative agreement is clearly found. %, at least for $F_1$.
A similar conclusion holds for $F_2$, given in Eqs. \eqref{summ2} and shown in Fig. \ref{Fig:10}.

Besides, the results of the evaluations of the rhombus and trapezoid diagrams in our approach are related, in the calculation scheme of Ref. \cite{Bali:2018nde}, to  lattice contractions called
$C_{1}$ and $C_{2}$. The correspondence formally reads
\begin{eqnarray}
C_{1}(q) & = & \frac{1}{2}X_{\Gamma}^{r}(q)\,,\nonumber \\
C_{2}(q) & = & \frac{1}{4}[X_{\Gamma}^{t}(q)+X_{\Gamma}^{p}(q)]\, .
\label{duerel}
\end{eqnarray}
The latter  quantities can also be directly compared to the findings
of Ref.~\cite{Bali:2018nde}, which will come of useful for the interpretation of contributions from diagrams.
Our results for these functions are shown in Figs.~\ref{Fig:8} and~\ref{Fig:9}, in coordinate and momentum space,
respectively. In particular the
former results 
show an encouraging agreement with the lattice data
reported, in coordinate space, in Figs. 11 (a) of Ref. \cite{Bali:2018nde}.
Overall, the NJL scheme seems to tentatively reproduce
the trend of the available lattice results.

Following an argument discussed in Ref.~\cite{Burkardt:1994pw}, it is observed that, in the absence of correlations,
two-current distributions could be computed from the single-current ones. This factorization hypothesis is comparable to an argument that has often
been used more recently for dPDFs
---see, {\it e.g.}, Ref.~\cite{Diehl1}.
This hypothesis is examined
in Ref.~\cite{Bali:2018nde}. In particular, for the isospin combination $\left\langle \pi^+\left(P\right)\right|\mathcal{O}_{n}^{uu}\left(y\right)\mathcal{O}_{n}^{dd}\left(0\right)\left|\pi^+\left(P\right)\right\rangle$, inserting a complete set of intermediate states between the two currents and keeping,
among them, only the pion ground state, it can be shown  that the two-current observable can be expressed in terms of single-current Form Factors.

In the most recent evaluation, the
the full lattice result 
for the $C_1$ contraction ---as it dominates the relevant isospin combination--- is compared to
single-current expressions using known results for the charges --{\it i.e.} Form Factors at zero momentum transfer--- as well as a common power law fall-off for the momentum transfer dependence of the Form Factor.
While the results at zero momentum transfer are acceptable, the attempt is found to generally fail to a large extent.

\begin{figure}[t]
\begin{center}
\includegraphics[scale=0.5]{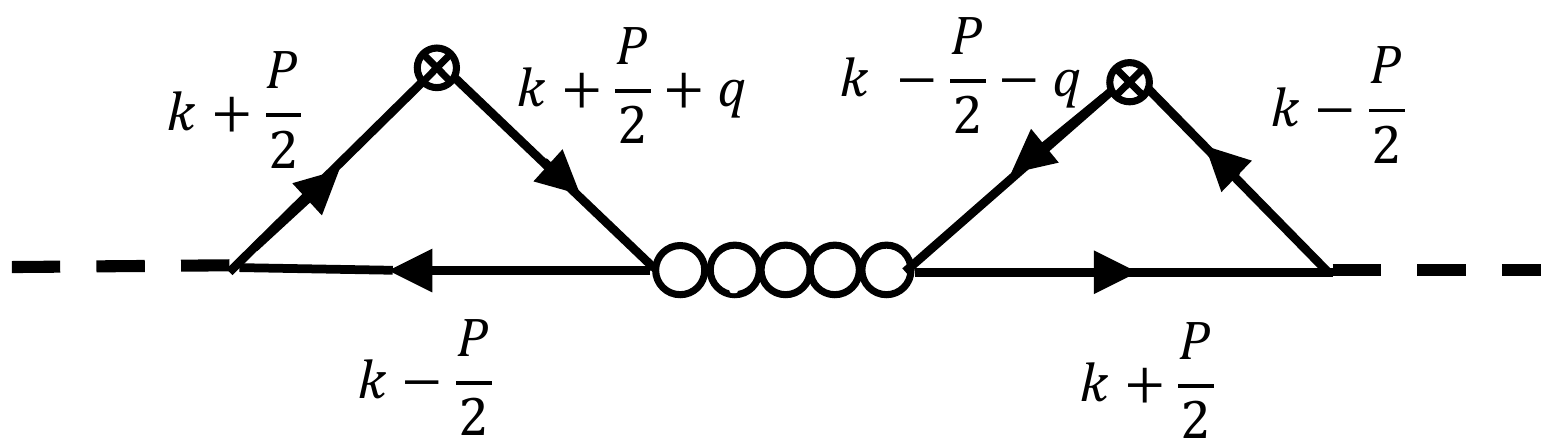}
\end{center}
\caption{One of the eight diagrams with pion intermediate states.}
\label{Fig_Two_Fem}
\end{figure}

Based on our previous experience with dPDFs~\cite{jhep19}, we have evaluated diagrams of the type shown in Fig.~\ref{Fig_Two_Fem},
where the intermediate  two  quark  state can have the quantum numbers of a $\sigma$ particle
or a pion, as allowed in the present calculation scheme.
After a tedious evaluation
detailed in the Appendix \ref{App.FormFactors}
for $\Gamma=\gamma^0$
and assuming dominance of the pion pole, in agreement with the procedure followed
in Ref.~\cite{Bali:2018nde},
it is found that the only
contribution arising
from these diagrams
can be written
\begin{equation}
X_{\Gamma}^{F_{PS}^{-}}(\vec{q})=4\,\frac{\left(E_{\pi}+m_{\pi}\right)^{2}}{E_{\pi}}\,\left[F_{em}\left(2m_{\pi}^{2}-2m_{\pi}E_{\pi}\right)\right]^{2} \, ,
\label{2ff}
\end{equation}
where $F_{em}$ is the pion electromagnetic form factor and $E_{\pi}=\sqrt{m_{\pi}^2+\vec{q}^2}$.
This expression ---after using the relevant isospin combination from Eq.~(\ref{eq:iso_pionexcha})--- agrees with that
adopted in Ref.~\cite{Bali:2018nde} to approximate the full result given by the function
$C_1$. In the NJL scheme used here
the contributions to $C_1$ arise from the {\it rhombus} diagram only;  
%does not receive 
such diagram cannot accommodate for 
any contribution of the type of Fig.~\ref{Fig_Two_Fem}. Therefore no comparison between the result of  
Eq.~(\ref{2ff}) with the contributions coming from $C_1$
---as suggested in Ref.~\cite{Bali:2018nde}---
 can be made according to the spirit of the present paper. Accordingly we do not discuss relations based on that factorization hypothesis any further.

\section{\label{V}Conclusions}

Two-partons correlations represent an elusive non perturbative information, theoretically encoded
in two-current correlations as well as double parton distribution functions.
The latter quantities are under theoretical investigation for the proton. There are  perspectives to observe
them during the LHC operation in the next years. While
a possible access to the same observable for the pion 
appears challenging, lattice results have already been 
obtained for two-current correlations in the pion,
a quantity that could be related to double parton distribution functions~\cite{Bali:2018nde}. A consistent field-theoretical approach, based
on the Nambu--Jona-Lasinio model with Pauli--Villars
regularization, has been used in this paper for
a systematic analysis 
of two-current
correlations in the pion.
We have given explicit expressions
for the time-time component of the vector-vector
two-current correlations in momentum space.
We have compared our results for
two-current correlations with lattice data, obtaining an overall good
qualitative agreement.
As happens in model calculation
of the related quantities double parton distributions, it is found  that, at the low scale relative to the model description, 
these functions encode novel non-perturbative information,
not present in one-body quantities. This conclusion is highlighted in the consideration of a possible factorization of the two-current correlation into one-body Form Factors, which is found to be an unconnected contribution in our low-energy model. 
It will be interesting to compare our results
for pion double distribution functions with forth-coming lattice data, whose
evaluation is presently in progress
\cite{private}.

\section*{Acknowledgments}

We thank M. Rinaldi for useful discussions and C. Zimmermann for sending us the lattice data of Ref. \cite{Bali:2018nde} relative to the quantities of interest here.
This work was supported in part by the Mineco
under contract FPA2016-77177-C2-1-P, {by the 
STRONG-2020 project of the European Union’s Horizon 2020 research and innovation programme under grant agreement No 824093,}  and by UNAM through the PIIF project Perspectivas en F\'isica de Part\'iculas y Astropart\'iculas as well as  Grant No. DGAPA-PAPIIT IA101720.
A.C. and S.S. thank the Department of Theoretical Physics of the
University of Valencia for warm hospitality and support.

\appendix

\section{The NJL model and regularization scheme}

\label{App.NJL_regularization}

\subsection{Basic physical quantities in the NJL model}
\label{App.NJL_Basic_Equations}

The Lagrangian density in the two-flavor version of the NJL model
is \cite{Klevansky:1992qe} 
\begin{equation}
\mathcal{L}=\bar{\psi}\left(i\not\partial-m_{0}\right)\psi+g\left[\left(\bar{\psi}\,\psi\right)^{2}+\left(\bar{\psi}\,\vec{\tau}\,i\gamma_{5}\,\psi\right)^{2}\right]\,\,,
\end{equation}
where $m_{0}$ is the current quark mass. The NJL is a chiral theory
that reproduces the spontaneous symmetry breaking process in which
the quark mass moves from the current value to its constituent value,
\begin{equation}
m=m_{0}-4g\,\left\langle \bar{u}u\right\rangle \,,
\end{equation}
where $\left\langle \bar{u}u\right\rangle $ is the quark condensate.

The main physical quantities associated to pion physics are defined
in terms of two integrals: 
\begin{equation}
I_{1}\left(m\right)=i\int\frac{d^{4}k}{\left(2\pi\right)^{4}}\frac{1}{k^{2}-m^{2}+i\epsilon}\label{I1.1}
\end{equation}
\begin{equation}
I_{2}\left(m,q^{2}\right)=i\int\frac{d^{4}k}{\left(2\pi\right)^{4}}\frac{1}{\left[\left(k+\frac{q}{2}\right)^{2}-m^{2}+i\epsilon\right]\left[\left(k-\frac{q}{2}\right)^{2}-m^{2}+i\epsilon\right]}\label{I2.1}
\end{equation}

Effectively, in the large $N_{c}$ approximation, the quark condensate
is defined by
\begin{equation}
\left\langle \bar{u}u\right\rangle =-4N_{c}mI_{1}\,.
\end{equation}
Pion and sigma masses are defined by the relations 
\begin{equation}
2g\,\Pi_{PS}\left(m_{\pi}^{2}\right)=1\,\,\,\,\,\,\,\,\,,\,\,\,\,\,\,\,\,\,\,\,\,\,2g\,\Pi_{S}\left(m_{\sigma}^{2}\right)=1\,,\label{NJL.17}
\end{equation}
with the scalar polarization 
\begin{align}
\Pi_{S}\left(q^{2}\right) & =-i\int\frac{d^{4}k}{\left(2\pi\right)^{4}}\text{Tr}\left[iS_{F}\left(p\right)\,iS_{F}\left(p-q\right)\right]\nonumber \\
 & =8N_{c}\left[I_{1}+\frac{1}{2}\left(4m^{2}-q^{2}\right)I_{2}\left(q\right)\right]\,,\label{NJL.14}
\end{align}
and the pseudoscalar polarization 
\begin{align}
\Pi_{PS}\left(q^{2}\right) & =-i\int\frac{d^{4}k}{\left(2\pi\right)^{4}}\text{Tr}\left[i\gamma_{5}\tau^{i}\,iS_{F}\left(p\right)i\gamma_{5}\tau^{i}\,iS_{F}\left(p-q\right)\right]\nonumber \\
 & =8N_{c}\left[I_{1}-\frac{1}{2}q^{2}I_{2}\left(q\right)\right]\,.\label{NJL.15}
\end{align}
The pion-quark and sigma-quark coupling constant are respectively
defined by 
\begin{align}
g_{\pi qq}^{2} & =\left(\frac{\partial\Pi_{PS}\left(q^{2}\right)}{\partial q^{2}}\right)_{q^{2}=m_{\pi}^{2}}^{-1}=\frac{-1}{4N_{c}\left[I_{2}\left(m_{\pi}^{2}\right)+m_{\pi}^{2}\left(\partial I_{2}/\partial q^{2}\right)_{q^{2}=m_{\pi}^{2}}\right]}\,,\nonumber \\
\label{NJL.20}\\
g_{\sigma qq}^{2} & =\left(\frac{\partial\Pi_{S}\left(q^{2}\right)}{\partial q^{2}}\right)_{q^{2}=m_{\sigma}^{2}}^{-1}=\frac{-1}{4N_{c}\left[I_{2}\left(m_{\sigma}^{2}\right)-\left(4m^{2}-m_{\sigma}^{2}\right)\left(\partial I_{2}/\partial q^{2}\right)_{q^{2}=m_{\sigma}^{2}}\right]}\,.\nonumber 
\end{align}
The pion decay constant is
\begin{equation}
f_{\pi}=-4N_{c}g_{\pi qq}mI_{2}\left(m_{\pi}^{2}\right)\,.
\end{equation}

The NJL model is a non-renormalizable field theory and a regularization
procedure has to be defined for the calculation of $I_{1}\left(m\right)$
and $I_{2}\left(m,q^{2}\right).$ We will introduce now the Pauli--Villars
regularization method for the NJL model.

\subsection{Pauli--Villars regularization scheme}
\label{App.NJL_PV_regularization}

In Section~\ref{II}, we have used the Pauli--Villars regularization
in order to render the occurring integrals finite. The way to proceed
in this method is: 
(1) remove from the numerator all the powers of the integrated momentum, which will be replaced by external momenta, and the mass of the constituent
quark, $m$; 
(2) for each resulting integral, which is of the form
\begin{equation}
\widetilde{I}_{n}\left(\mu\left(m\right)\right)=\int\frac{d^{4}k}{\left(2\pi\right)^{4}}\frac{1}{\left[k^{2}-\mu\left(m\right)^{2}+i\epsilon\right]^{n}}\,,
\end{equation}
make the substitution
\begin{equation}
\widetilde{I}_{n}^{r}\left(\mu\left(m\right)\right)=\sum_{j=0}^{2}\,c_{j}\,\widetilde{I}_{n}\left(\mu\left(M_{j}\right)\right)\,,
\end{equation}
with $M_{j}^{2}=m^{2}+j\,\Lambda^{2}$, $c_{0}=c_{2}=1$ and $c_{1}=-2$.

Following this procedure, we obtain for the momentum integral of one propagator
\begin{equation}
I_{1}=\frac{1}{16\pi^{2}}\sum_{j=0}^{2}c_{j}M_{j}^{2}\ln\frac{M_{j}^{2}}{m^{2}}\,,
\end{equation}
and for the one of two propagators  
\begin{align}
I_{2}\left(m,q^{2}\right) & =\frac{1}{16\pi^{2}}\sum_{j=0}^{2}c_{j}\left(\ln\frac{M_{j}^{2}}{m^{2}}+2\sqrt{\frac{4M_{j}^{2}}{q^{2}}-1}\,\arctan\frac{1}{\sqrt{\frac{4M_{j}^{2}}{q^{2}}-1}}\right)\,.
\end{align}
With the conventional values $\left\langle \bar{u}u\right\rangle =-(0.250\,\text{GeV})^{3},$
$f_{\pi}=0.0924\,\text{GeV}$ and $m_{\pi}=0.140\,\text{GeV}$, we
get $m=0.238\,\text{GeV}$, $\Lambda$=0.860 GeV and $m_{0}=5.4\,\text{MeV}.$
For the pion-quark coupling constant we get $g_{\pi qq}^{2}=6.279.$
We can obtain the chiral limit taking $m_{0}=0$, without changing
$\Lambda$ and $m.$ In that case $\left\langle \bar{u}u\right\rangle $
and $f_{\pi}$ do not change but one has $m_{\pi}=0$ and $g_{\pi qq}^{2}=6.625.$

For a proper comparison with the lattice data of Ref. \cite{Bali:2018nde},
we have applied the same model with $\Lambda=1.022\,\text{GeV,}$
$m=0.242\,\text{GeV}$ and $m_{0}=20\,\text{MeV}$, which lead to a
massive pion with $m_{\pi}=0.300\,\text{GeV,}$ $f_{\pi}=0.1\,\text{GeV}$
and $\left\langle \bar{u}u\right\rangle =-(0.285\,\text{GeV})^{3}$,
as used in Ref. \cite{Bali:2018nde}. In this case we have $g_{\pi qq}^{2}=4.527.$

\subsection{Intermediate pion states}
\label{App.FormFactors}

Let us consider now 
the evaluation of the calculation of the process depicted in Fig. \ref{Fig_Two_Fem}.
There are eight different diagrams
of this type.

This contribution can be written in the following form
\begin{align}
G_{\alpha\beta}^{\left(F\right)ij}\left(\vec{q}\right)= & \int\frac{dq^{0}}{2\pi}\left[\mathbb{U}_{j\beta}\frac{2ig}{1-2g\,\Pi_{S}\left(\left(P+q\right)^{2}\right)}\mathbb{V}_{i\alpha}+\left(q^{\mu}\leftrightarrow-q^{\mu},\,\alpha\leftrightarrow\beta\right)\right]\nonumber \\
+ & \int\frac{dq^{0}}{2\pi}\left[\sum_{c=1}^{3}\mathbb{\widetilde{U}}_{j\beta c}\frac{2ig}{1-2g\,\Pi_{PS}\left(\left(P+q\right)^{2}\right)}\mathbb{\widetilde{V}}_{i\alpha c}+\left(q^{\mu}\leftrightarrow-q^{\mu},\,\alpha\leftrightarrow\beta\right)\right]\label{DosFem1} \, .
\end{align}
The first (second) line of this equation describes an intermediate
two quark state with the quantum numbers of a $\sigma(\pi)$ particle.
In the case of the intermediate pion the additional isospin index,
$c,$ corresponds to its isospin. The quantities
introduced in Eq.~(\ref{DosFem1}) are
\begin{align}
\mathbb{U}_{j\beta} & =-\int\frac{d^{4}k}{\left(2\pi\right)^{4}}\text{Tr}\left[\bar{\phi}_{\pi^{j}}iS_{F}\left(k+\frac{P}{2}\right)\tau^{\beta}\Gamma iS_{F}\left(k+\frac{P}{2}+q\right)iS_{F}\left(k-\frac{P}{2}\right)\right] \nonumber \\
 & -\int\frac{d^{4}k}{\left(2\pi\right)^{4}}\text{Tr}\left[\bar{\phi}_{\pi^{j}}iS_{F}\left(k+\frac{P}{2}\right)iS_{F}\left(k-\frac{P}{2}-q\right)\tau^{\beta}\Gamma iS_{F}\left(k-\frac{P}{2}\right)\right] \, ,
\end{align}
\begin{align}
\mathbb{V}_{i\alpha} & =-\int\frac{d^{4}k}{\left(2\pi\right)^{4}}\text{Tr}\left[\phi_{\pi^{i}}iS_{F}\left(k-\frac{P}{2}\right)iS_{F}\left(k+\frac{P}{2}+q\right)\tau^{\alpha}\Gamma iS_{F}\left(k+\frac{P}{2}\right)\right]\nonumber \\
 & -\int\frac{d^{4}k}{\left(2\pi\right)^{4}}\text{Tr}\left[\phi_{\pi^{i}}iS_{F}\left(k-\frac{P}{2}\right)\tau^{\alpha}\Gamma iS_{F}\left(k-\frac{P}{2}-q\right)iS_{F}\left(k+\frac{P}{2}\right)\right]
 \, ,
\end{align}
\begin{align}
\mathbb{\widetilde{U}}_{j\beta c} & =-\int\frac{d^{4}k}{\left(2\pi\right)^{4}}\text{Tr}\left[\bar{\phi}_{\pi^{j}}iS_{F}\left(k+\frac{P}{2}\right)\tau^{\beta}\Gamma iS_{F}\left(k+\frac{P}{2}+q\right)i\gamma_{5}\tau^{c}iS_{F}\left(k-\frac{P}{2}\right)\right]\nonumber \\
 & -\int\frac{d^{4}k}{\left(2\pi\right)^{4}}\text{Tr}\left[\bar{\phi}_{\pi^{j}}iS_{F}\left(k+\frac{P}{2}\right)i\gamma_{5}\tau^{c}iS_{F}\left(k-\frac{P}{2}-q\right)\tau^{\beta}\Gamma iS_{F}\left(k-\frac{P}{2}\right)\right]
\, ,
\end{align}
\begin{align}
\mathbb{\widetilde{V}}_{i\alpha c} & =-\int\frac{d^{4}k}{\left(2\pi\right)^{4}}\text{Tr}\left[\phi_{\pi^{i}}iS_{F}\left(k-\frac{P}{2}\right)i\gamma_{5}\tau^{c}iS_{F}\left(k+\frac{P}{2}+q\right)\tau^{\alpha}\Gamma iS_{F}\left(k+\frac{P}{2}\right)\right]\nonumber \\
 & -\int\frac{d^{4}k}{\left(2\pi\right)^{4}}\text{Tr}\left[\phi_{\pi^{i}}iS_{F}\left(k-\frac{P}{2}\right)\tau^{\alpha}\Gamma iS_{F}\left(k-\frac{P}{2}-q\right)i\gamma_{5}\tau^{c}iS_{F}\left(k+\frac{P}{2}\right)\right]
\, .
\end{align}
Performing the isospin traces, we obtain the following structures
(we disregard the $G_{a\text{s}}^{\left(F\right)ij}$ and $G_{\text{s}a}^{\left(F\right)ij}$
cases):
\begin{eqnarray}
G_{\text{ss}}^{\left(F\right)ij}\left(\vec{q}\right) & = &-\delta_{ij}\,X_{\Gamma}^{F_{PS}^{+}}\left(\vec{q}\right) \, ,
\nonumber \\
G_{ab}^{\left(F\right)ij}\left(\vec{q}\right) & = & -\delta_{ij}\,\delta_{ab}\,X_{\Gamma}^{F_{PS}^{-}}\left(\vec{q}\right)+
\left(\delta_{ja}\delta_{ib}+\delta_{jb}\delta_{ia}\right)\frac{1}{2}\left[X_{\Gamma}^{F_{S}}\left(\vec{q}\right)+X_{\Gamma}^{F_{PS}^{-}}\left(\vec{q}\right)\right] \, , \label{DosFem4}\\
 & + & \left(\delta_{ja}\delta_{ib}-\delta_{jb}\delta_{ia}\right)\frac{1}{2}\left[Z_{\Gamma}^{F_{S}}\left(\vec{q}\right)-Z_{\Gamma}^{F_{PS}^{-}}\left(\vec{q}\right)\right] \, ,
\nonumber 
\label{eq:iso_pionexcha}
\end{eqnarray}
with
\begin{align}
X_{\Gamma}^{F_{S}}\left(\vec{q}\right) & =\left(ig_{\pi qq}\right)^{2}\,4N_{c}^{2}\int\frac{dq^{0}}{2\pi}\,\mathbb{H}_{\Gamma}^{\textrm{U}^{+}}\left(q\right)\,
\frac{2ig}{1-2g\,\Pi_{S}\left(\left(P+q\right)^{2}\right)}\,\mathbb{H}_{\Gamma}^{\textrm{V}^{+}}
\left(q\right)+\left(\vec{q}\leftrightarrow-\vec{q}\right)\, , \nonumber \\
X_{\Gamma}^{F_{PS}^{+}}\left(\vec{q}\right) & =\left(ig_{\pi qq}\right)^{2}\,4N_{c}^{2}\int\frac{dq^{0}}{2\pi}\,\mathbb{H}_{\Gamma}^{\widetilde{\textrm{U}}^{+}}\left(q\right)\,
\frac{2ig}{1-2g\,\Pi_{PS}\left(\left(P+q\right)^{2}\right)}\,\mathbb{H}_{\Gamma}^{\widetilde{\textrm{V}}^{+}}\left(q\right)+
\left(\vec{q}\leftrightarrow-\vec{q}\right)\label{XZ_DosFem1}\, , \\
X_{\Gamma}^{F_{PS}^{-}}(\vec{q}) & =\left(ig_{\pi qq}\right)^{2}\,4N_{c}^{2}\int\frac{dq^{0}}{2\pi}\,
\mathbb{H}_{\Gamma}^{\widetilde{\textrm{U}}^{-}}\left(q\right)\,\frac{2ig}{1-2g\,\Pi_{PS}\left(\left(P+q\right)^{2}\right)}\,
\mathbb{H}_{\Gamma}^{\widetilde{\textrm{V}}^{-}}\left(q\right)+\left(\vec{q}\leftrightarrow-\vec{q}\right)\, , \nonumber 
\end{align}
\begin{align*}
Z_{\Gamma}^{F_{S}}\left(\vec{q}\right) & =\left(ig_{\pi qq}\right)^{2}\,4N_{c}^{2}\int\frac{dq^{0}}{2\pi}\,\mathbb{H}_{\Gamma}^{\textrm{U}^{+}}\left(q\right)
\,\frac{2ig}{1-2g\,\Pi_{S}\left(\left(P+q\right)^{2}\right)}\,\mathbb{H}_{\Gamma}^{\textrm{V}^{+}}
\left(q\right)-\left(\vec{q}\leftrightarrow-\vec{q}\right)\, , \\
Z_{\Gamma}^{F_{PS}^{-}}(\vec{q}) & =\left(ig_{\pi qq}\right)^{2}\,4N_{c}^{2}\int
\frac{dq^{0}}{2\pi}\,\mathbb{H}_{\Gamma}^{\widetilde{\textrm{U}}^{-}}\left(q\right)\,\frac{2ig}{1-2g\,\Pi_{PS}
\left(\left(P+q\right)^{2}\right)}\,\mathbb{H}_{\Gamma}^{\widetilde{\textrm{V}}^{-}}
\left(q\right)-\left(\vec{q}\leftrightarrow-\vec{q}\right) \, ,
\end{align*}
where
\begin{align}
\mathbb{H}_{\Gamma}^{\ell^{\pm}}\left(q\right) & =\int\frac{d^{4}k}{\left(2\pi\right)^{4}}\,\frac{1}{\left[\left(k+\frac{P}{2}\right)^{2}-m^{2}+i\epsilon\right]\left[\left(k-\frac{P}{2}\right)^{2}-m^{2}+i\epsilon\right]} \nonumber \\
 & \left(\frac{t_{\Gamma}^{\ell_{1}}\left(k\right)}{\left[\left(k+\frac{P}{2}+q\right)^{2}-m^{2}+i\epsilon\right]}\pm\frac{t_{\Gamma}^{\ell_{2}}\left(k\right)}{\left[\left(k-\frac{P}{2}-q\right)^{2}-m^{2}+i\epsilon\right]}\right) \, ,\label{DosFemH}
\end{align}
\begin{align}
t_{\Gamma}^{\text{U}_{1}} & =\text{tr}\left[\gamma_{5}\left(\cancel{k}+\frac{\cancel{P}}{2}+m\right)\Gamma\left(\cancel{k}+\frac{\cancel{P}}{2}+\cancel{q}+m\right)\left(\cancel{k}-\frac{\cancel{P}}{2}+m\right)\right]\, , \nonumber \\
t_{\Gamma}^{\text{U}_{2}} & =\text{tr}\left[\gamma_{5}\left(\cancel{k}+\frac{\cancel{P}}{2}+m\right)\left(\cancel{k}-\frac{\cancel{P}}{2}-\cancel{q}+m\right)\Gamma\left(\cancel{k}-\frac{\cancel{P}}{2}+m\right)\right]\, , \nonumber \\
t_{\Gamma}^{\text{V}_{1}} & =\text{tr}\left[\gamma_{5}\left(\cancel{k}-\frac{\cancel{P}}{2}+m\right)\left(\cancel{k}+\frac{\cancel{P}}{2}+\cancel{q}+m\right)\Gamma\left(\cancel{k}+\frac{\cancel{P}}{2}+m\right)\right]\, , \\
t_{\Gamma}^{\text{V}_{2}} & =\text{tr}\left[\gamma_{5}\left(\cancel{k}-\frac{\cancel{P}}{2}+m\right)\Gamma\left(\cancel{k}-\frac{\cancel{P}}{2}-\cancel{q}+m\right)\left(\cancel{k}+\frac{\cancel{P}}{2}+m\right)\right]\, , \nonumber 
\end{align}
\begin{align}
t_{\Gamma}^{\widetilde{\text{U}}_{1}} & =\text{tr}\left[\gamma_{5}\left(\cancel{k}+\frac{\cancel{P}}{2}+m\right)\Gamma\left(\cancel{k}+\frac{\cancel{P}}{2}+\cancel{q}+m\right)\gamma_{5}\left(\cancel{k}-\frac{\cancel{P}}{2}+m\right)\right]\, , \nonumber \\
t_{\Gamma}^{\widetilde{\text{U}}_{2}} & =\text{tr}\left[\gamma_{5}\left(\cancel{k}+\frac{\cancel{P}}{2}+m\right)\gamma_{5}\left(\cancel{k}-\frac{\cancel{P}}{2}-\cancel{q}+m\right)\Gamma\left(\cancel{k}-\frac{\cancel{P}}{2}+m\right)\right]\, , \nonumber \\
t_{\Gamma}^{\widetilde{\text{V}}_{1}} & =\text{tr}\left[\gamma_{5}\left(\cancel{k}-\frac{\cancel{P}}{2}+m\right)\gamma_{5}\left(\cancel{k}+\frac{\cancel{P}}{2}+\cancel{q}+m\right)\Gamma\left(\cancel{k}+\frac{\cancel{P}}{2}+m\right)\right]\, , \\
t_{\Gamma}^{\widetilde{\text{V}}_{2}} & =\text{tr}\left[\gamma_{5}\left(\cancel{k}-\frac{\cancel{P}}{2}+m\right)\Gamma\left(\cancel{k}-\frac{\cancel{P}}{2}-\cancel{q}+m\right)\gamma_{5}\left(\cancel{k}+\frac{\cancel{P}}{2}+m\right)\right]\, . \nonumber 
\end{align}

Using Eqs.~(\ref{NJL.17}), (\ref{NJL.14}), (\ref{NJL.15}) and (\ref{NJL.20}),
the intermediate two quark states appearing in Eq.~(\ref{XZ_DosFem1})
can be approximated by the quantities, 
\begin{align}
\frac{2ig}{1-2g\,\Pi_{S}\left(\left(P+q\right)^{2}\right)} & \simeq\frac{-ig_{\sigma qq}^{2}}{\left(P+q\right)^{2}-m_{\sigma}^{2}+i\epsilon}
\, , \nonumber \\
\frac{2ig}{1-2g\,\Pi_{PS}\left(\left(P+q\right)^{2}\right)} & \simeq\frac{-ig_{\pi qq}^{2}}{\left(P+q\right)^{2}-m_{\pi}^{2}+i\epsilon}
\, .\label{MesonProp}
\end{align}

Let us consider, as always in this paper, the $\Gamma=\gamma^{0}$ case. Performing the traces
present in Eq.~(\ref{DosFemH}) we have
\begin{equation}
\mathbb{H}_{V^{0}}^{\textrm{U}^{+}}\left(q\right)=\mathbb{H}_{V^{0}}^{\textrm{V}^{+}}\left(q\right)=\mathbb{H}_{V^{0}}^{\widetilde{\textrm{U}}^{+}}\left(q\right)=\mathbb{H}_{V^{0}}^{\widetilde{\textrm{V}}^{+}}\left(q\right)=0\,,
\end{equation}
so that $X_{V^{0}}^{F_{S}}\left(\vec{q}\right)=X_{V^{0}}^{F_{PS}^{+}}\left(\vec{q}\right)=Z_{\Gamma}^{F_{S}}\left(\vec{q}\right)=0.$
After a tedious but straightforward calculation we obtain,
\begin{equation}
\mathbb{H}_{\Gamma}^{\widetilde{\textrm{U}}^{-}}\left(q\right)=\mathbb{H}_{\Gamma}^{\widetilde{\textrm{V}}^{-}}\left(q\right)=\frac{1}{N_{c}g_{\pi qq}^{2}}\left[\left(2P+q\right)^{0}F_{em}\left(q^{2}\right)+q^{0}\,\widetilde{F}_{em}\left(q^{2}\right)\right] \, ,
\end{equation}
where $F_{em}\left(q^{2}\right)$ and $\widetilde{F}_{em}\left(q^{2}\right)$
are
\begin{eqnarray}
F_{em}\left(q^{2}\right) & = & \frac{2N_{c}g_{\pi qq}^{2}}{P^{2}P^{\prime2}-\left(P\cdot P^{\prime}\right)^{2}}\left[P^{2}\left(P\cdot P^{\prime}-P^{\prime2}\right)I_{2}\left(P^{2}\right)+P^{\prime2}\left(P\cdot P^{\prime}-P^{2}\right)I_{2}\left(P^{\prime2}\right)\right.\nonumber \\
 & + & \left. \left(P\cdot P^{\prime}\right)\,\left(2\,P\cdot P^{\prime}-P^{2}-P^{\prime2}\right)I_{2}\left(\left(P-P^{\prime}\right)^{2}\right)
 \right.\nonumber \\ 
& - & \left . P^{2}P^{\prime2}\left(2\,P\cdot P^{\prime}-P^{2}-P^{\prime2}\right)I_{3}\left(P,P^{\prime}\right)\right]\, ,
\end{eqnarray}
\begin{align}
\widetilde{F}_{em}\left(q^{2}\right) & =\frac{2N_{c}g_{\pi qq}^{2}}{P^{2}P^{\prime2}-\left(P\cdot P^{\prime}\right)^{2}}\left[P^{2}\left(P\cdot P^{\prime}-P^{\prime2}\right)I_{2}\left(P^{2}\right)-P^{\prime2}\left(P\cdot P^{\prime}-P^{2}\right)I_{2}\left(P^{\prime2}\right)\right.\nonumber \\
 & \left.+\left(P\cdot P^{\prime}\right)\,\left(P^{\prime2}-P^{2}\right)I_{2}\left(\left(P-P^{\prime}\right)^{2}\right)-P^{2}P^{\prime2}\left(P^{\prime2}-P^{2}\right)I_{3}\left(P,P^{\prime}\right)\right] \, ,
\end{align}
with $P^{\prime\mu}=P^{\mu}+q^{\mu}.$ These two form factors correspond
to the standard electromagnetic form factors,
\begin{equation}
\left\langle \pi\left(P\right)\right|J_{em}^{\mu}\left|\pi\left(P^{\prime}\right)\right\rangle =\left(P^{\prime}+P\right)^{\mu}F_{em}\left(q^{2}\right)+\left(P^{\prime}-P\right)^{\mu}\widetilde{F}_{em}\left(q^{2}\right)\,.
\end{equation}
When both pions are on-shell, $\widetilde{F}_{em}\left(q^{2}\right)$ vanishes  and $F_{em}\left(q^{2}\right)$ reads
\begin{equation}
F_{em}\left(q^{2}\right)=\frac{4N_{c}g_{\pi qq}^{2}}{m_{\pi}^{2}+\left(P\cdot P^{\prime}\right)}\left[-m_{\pi}^{2}I_{2}\left(m_{\pi}^{2}\right)-\left(P\cdot P^{\prime}\right)\,I_{2}\left(\left(P-P^{\prime}\right)^{2}\right)+m_{\pi}^{4}I_{3}\left(P,P^{\prime}\right)\right]
\end{equation}
\\

Putting together all these ingredients we have
\begin{align}
X_{\Gamma}^{F_{PS}^{-}}(\vec{q}) & =i\,8\int\frac{dq^{0}}{2\pi}\,\left[\left(2P+q\right)^{0}F_{em}\left(q^{2}\right)+q^{0}\,\widetilde{F}_{em}\left(q^{2}\right)\right]^{2}\,\frac{1}{\left(P+q\right)^{2}-m_{\pi}^{2}+i\epsilon}\, , \nonumber \\
Z_{\Gamma}^{F_{PS}^{-}}(\vec{q}) & =0 \, .
\end{align}
The $q^{0}$ integral will be dominated by the pole contribution and we obtain the final result
\begin{equation}
X_{\Gamma}^{F_{PS}^{-}}(\vec{q})=4\,\frac{\left(E_{\pi}+m_{\pi}\right)^{2}}{E_{\pi}}\,\left[F_{em}\left(2m_{\pi}^{2}-2m_{\pi}E_{\pi}\right)\right]^{2} \, ,
\end{equation}
which is the expression reported at the end of Section~\ref{IV}.

\end{document}